\documentclass[a4paper, 11pt, oneside]{article}
\pdfoutput=1
\usepackage{float}
\usepackage{jheppub}
\usepackage{epstopdf}
\usepackage{color}
\usepackage[utf8]{inputenc}

\usepackage{bbm}   
\usepackage{amsmath}
\usepackage{graphicx}

\newcommand{\nc}{\newcommand}

\newlength{\absize}
\nc{\non}{\nonumber}
\nc{\hc}{\hbox {H.c.}} 
\nc{\noi}{\noindent}
\nc{\barx}{\bar{x}}
\nc{\pbarn}{\;\hbox {pb}}
\nc{\fbarn}{\;\hbox {fb}}
\newcommand{\bi}{\begin{itemize}}
\newcommand{\ei}{\end{itemize}}

\def\thetaW{{\theta}_\text{W}}
\nc{\lsp}{\;\;\;\;\;}
\nc{\Lsp}{\;\;\;\;\;\;\;\;\;\;}  
\nc{\LLsp}{\lspace \lspace}
\nc{\lra}{\longrightarrow}

\def\thetaW{{\theta}_\text{W}}

\newcommand{\half}{{\textstyle\frac{1}{2}}}
\newcommand{\threehalf}{{\textstyle\frac{3}{2}}}

\nc{\beq}{\begin{equation}}  \nc{\eeq}{\end{equation}}
\nc{\bea}{\begin{eqnarray}}  \nc{\eea}{\end{eqnarray}}
\nc{\baa}{\begin{array}}     \nc{\eaa}{\end{array}}
\nc{\bit}{\begin{itemize}}   \nc{\eit}{\end{itemize}}
\nc{\ben}{\begin{enumerate}} \nc{\een}{\end{enumerate}}
\nc{\bce}{\begin{center}}    \nc{\ece}{\end{center}}
\nc{\bpm}{\begin{pmatrix}}   \nc{\epm}{\end{pmatrix}}
\nc{\bvt}{\begin{verbatim}}  \nc{\evt}{\end{verbatim}}
\def\lsim{\mathrel{\raise.3ex\hbox{$<$\kern-.75em\lower1ex\hbox{$\sim$}}}}
\def\gsim{\mathrel{\raise.3ex\hbox{$>$\kern-.75em\lower1ex\hbox{$\sim$}}}}

\def\mcal{{\cal M}}

\nc{\tanb}{\tan\beta}
\nc{\mch}{M_{H^\pm}}

\def\mch{M_{H^\pm}}

\nc{\for}{\lsp {\rm for} \lsp}
\nc{\andd}{\lsp {\rm and} \lsp}

\renewcommand{\Re}{\mbox{Re\thinspace}}
\renewcommand{\Im}{\mbox{Im\thinspace}}

\allowdisplaybreaks 

\def\i11{{\mathbbm 1}}

\title{CP-Violation in the \boldmath{$ZZZ$} and \boldmath{$ZWW$} vertices at \boldmath{$e^+e^-$} colliders in Two-Higgs-Doublet Models}
\author[a]{B. Grzadkowski,}
\affiliation[a]{Faculty of Physics, University of Warsaw, Pastura 5, 02-093 Warsaw, Poland}
\author[b]{O. M. Ogreid,}
\affiliation[b]{Bergen University College, Postboks 7030, N-5020 Bergen, Norway}
\author[c]{P. Osland}
\affiliation[c]{Department of Physics,
University of Bergen, Postboks 7803, N-5020 Bergen, Norway}
\emailAdd{bohdan.grzadkowski@fuw.edu.pl}
\emailAdd{omo@hib.no}
\emailAdd{Per.Osland@ift.uib.no}
\date{\today}

\abstract{We discuss possibilities of measuring CP violation in the Two-Higgs-Doublet Model by studying effects of
one-loop generated $ZZZ$ and $ZWW$ vertices. We discuss a set of CP-sensitive asymmetries
for $ZZ$ and $W^+W^-$ production at linear $e^+e^-$-colliders, that directly depends on the weak-basis invariant $\Im J_2$ that parametrises the strength of CP violation. 
Given the restrictions on this model that follow from the LHC measurements, the predicted effects are small. Pursuing such measurements is however very important, as an observed signal might point to a richer scalar sector.
}

\begin{document}
\maketitle

\flushbottom

\section{Introduction}
\label{Sec:Introduction}

Anomalous contributions to trilinear electroweak vector boson couplings have been thoroughly studied 
\cite{Gaemers:1978hg,Hagiwara:1986vm,Gounaris:1999kf,Gounaris:2000dn,Baur:2000ae} and searched for, 
at LEP \cite{Schael:2013ita}, at Fermilab \cite{Aaltonen:2008mv,Aaltonen:2009fd,Abazov:2011td,Abazov:2012cj} 
and at the LHC \cite{Aad:2011xj,Chatrchyan:2012sga,Aad:2012awa,Edwards:2013,Gregersen:2013lza,ATLAS:2013gma,Chatrchyan:2013yaa,CMS:2014xja,Aad:2014mda,Khachatryan:2015pba,Khachatryan:2015sga}.
Experimentally, the 
\begin{equation}
VW^+W^-, \quad V=\gamma,Z
\end{equation}
couplings are considered the more accessible, whereas the
\begin{equation}
VZZ, \quad V=\gamma,Z
\end{equation}
couplings are considered more challenging. Both classes may have a CP-violating, as well as a CP-conserving part.

In the Standard Model (SM), at the tree level, only the $\gamma WW$ and $ZWW$ couplings are non-zero, whereas all four receive contributions at the one-loop level. In the SM, CP-violating effects can only be induced via the CKM matrix. However, at one-loop order, there is no such contribution, since there might be only two relatively complex-conjugated $\bar{q} q^\prime W$ vertices, hence CP-violating phases of the CKM matrix would cancel. An extended Higgs sector may naturally modify this at the one-loop level, since new sources of CP violation could enter in a non-trivial way.

As is well known, the Two-Higgs-Doublet Model allows for CP violation, either explicit or spontaneous 
 \cite{Lee:1973iz}. Early work on CP violation in the Higgs sector related it to the couplings of neutral scalars to the electroweak gauge bosons, as well as to the charged scalars \cite{Lavoura:1994fv,Botella:1994cs}.
  The conditions for having CP violation in the model can be expressed in terms of three invariants, in Ref.~\cite{Gunion:2005ja} denoted $\Im J_1$, $\Im J_2$ and $\Im J_3$. If any one of them is non-zero, then CP is violated \cite{Gunion:2005ja} (see also Ref.~\cite{Branco:2005em}). Further criteria would allow to distinguish spontaneous and explicit CP violation \cite{Davidson:2005cw,Gunion:2005ja}.

Standard-model contributions to the $ZZZ$ and $ZWW$ vertices have been studied in \cite{Denner:1988tv} and \cite{Bohm:1987ck}, respectively. Since there is some scope for further constraining or even measuring CP violation in these couplings, we present an updated review of these observables, and also propose some new ones.

The paper is organized as follows. After a brief review of the model and the basic CP-violating invariants in section~2, we discuss one-loop contributions to the $ZZZ$ and $ZWW$ vertices in sections~3 and 4. Selected CP-violating asymmetries that could be measured in $e^+e^-$ collisions are discussed in section~5, and concluding remarks are given in section~6. Technical details are relegated to appendices.
\section{The model}
\label{Sec:def-model}
\setcounter{equation}{0}

We adopt a standard parametrization for the scalar potential of the 2HDM (see, for example, \cite{Grzadkowski:2014ada}) 
with
\begin{equation}
\Phi_i=\left(
\begin{array}{c}\varphi_i^+\\ (v_i+\eta_i+i\chi_i)/\sqrt{2}
\end{array}\right), \quad
i=1,2.
\label{Eq:basis}
\end{equation}
In the general CP-violating case, the model contains three neutral scalars, which are linear compositions of the $\eta_i$ and $\chi_i$:
\begin{equation} \label{Eq:R-def}
\begin{pmatrix}
H_1 \\ H_2 \\ H_3
\end{pmatrix}
=R
\begin{pmatrix}
\eta_1 \\ \eta_2 \\ \eta_3
\end{pmatrix},
\end{equation}
with $\eta_3$ a linear combination of the $\chi_i$ that is
orthogonal to the Goldstone field $G_0$.
Furthermore, the $3\times3$ rotation matrix $R$ satisfies
\begin{equation}
\label{Eq:cal-M}
R{\cal M}^2R^{\rm T}={\cal M}^2_{\rm diag}={\rm diag}(M_1^2,M_2^2,M_3^2),
\end{equation}
where ${\cal M}^2$ is the neutral-sector mass-squared matrix,
and with $M_1\leq M_2\leq M_3$.

The weak-basis invariants revealing CP violation were originally expressed by Lavoura, Silva and Botella  \cite{Lavoura:1994fv,Botella:1994cs}, in terms of couplings and rotation-matrix elements.
The notation $\Im J_i$, where the invariants were expressed in terms of potential parameters was introduced by Gunion and Haber \cite{Gunion:2005ja}. It was recently discussed in more detail by the present authors \cite{Grzadkowski:2014ada} (where also $\Im J_3$ was replaced by another related invariant which we named $\Im J_{30}$). The invariant $\Im J_2$, which represents CP violation in the mass matrix, can be written as
\begin{align} \label{Eq:Im_J2}
\Im J_2&=\frac{2e_1 e_2 e_3}{v^9}(M_1^2-M_2^2)(M_2^2-M_3^2)(M_3^2-M_1^2)\nonumber \\
&\quad =\frac{2e_1 e_2 e_3}{v^9}\sum_{i,j,k}\epsilon_{ijk}M_i^4M_k^2,
\end{align}
where $M_i$ are the neutral Higgs masses, and $e_i\equiv v_1R_{i1}+v_2R_{i2}$ represents their couplings to a $Z$ or a $W$ (for a full dictionary of couplings determined by $e_i$, see appendix~B of Ref.~\cite{Grzadkowski:2014ada}).

We shall in this paper focus on processes in which $\Im J_2$ is responsible for the CP violation. This invariant is the only one which does not involve charged scalars. Charged scalars are involved in processes for which $\Im J_1$ and/or $\Im J_{30}$ are responsible for the CP violation. For the explicit form of these invariants and processes to which they contribute, we refer to  Ref.~\cite{Grzadkowski:2014ada}. 

\section{The \boldmath{$ZZZ$} vertex}
\label{Sec:ZZZ}
\setcounter{equation}{0}

\begin{figure}[htb]
	\vspace*{-2.0cm}
	\centerline{
		\includegraphics[width=15.0cm]{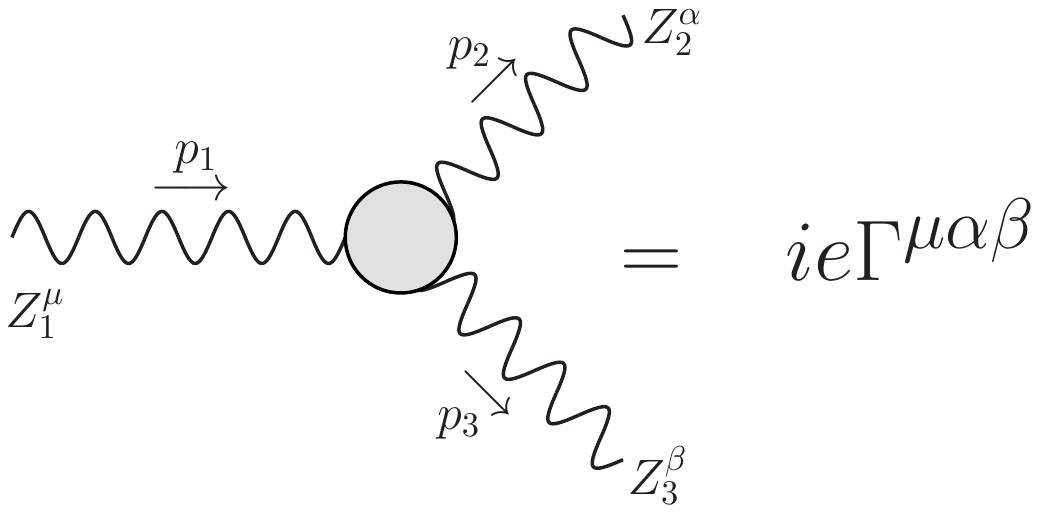}}
	\vspace*{-16.0cm}
	\caption{The general $ZZZ$ vertex.}
	\label{fig:blob-zzz}
	
\end{figure}

One of the simplest vertex functions to which $\Im J_2$ contributes, is the effective $ZZZ$ vertex discussed in appendix~\ref{Sec:zzz}. Since each $ZH_iH_j$ vertex contains a factor $\epsilon_{ijk}$ (see appendix~B of 
Ref.~\cite{Grzadkowski:2014ada}), it follows that $i,j,k$ must be some permutation of $1,2,3$ and thus an over-all factor of $e_1e_2e_3$ will emerge.

CP-violating form factors for triple gauge boson couplings have previously been studied in the 2HDM in Refs.~\cite{He:1992qh,Chang:1994cs,Chang:1993vv}.
\subsection{Lorentz structure}
Phenomenological discussions \cite{Hagiwara:1986vm,Gounaris:1999kf,Gounaris:2000dn,Baur:2000ae} of the $ZZZ$ vertex have presented its 
most general Lorentz structure.
In Ref.~\cite{Gounaris:2000dn} the CP-violating vertex is analyzed, with all $Z_1, Z_2, Z_3$ off-shell. A total of 14 Lorentz structures are identified, all preserving parity. Some of these vanish when one or more $Z$ is on-shell. 
(For a detailed discussion of this structure, see Ref.~\cite{Nieves:1996ff}.)
We characterize them by momenta and 
Lorentz indices ($p_1,\mu$), ($p_2,\alpha$) and ($p_3,\beta$), and let $Z_1$ be off-shell while $Z_2$ and $Z_3$ are on-shell. In addition, we assume that $Z_1$ couples to a pair of leptons $e^+e^-$ and neglect terms proportional to the lepton mass. Then 
according to \cite{Gounaris:1999kf} the structure reduces to the form\footnote{Here, we follow the convention of Hagiwara et al \cite{Hagiwara:1986vm}, which we also adopt in section~\ref{Sec:ZWW} for the $ZWW$ vertex by putting $\epsilon_{0123}=-\epsilon^{0123}=+1$, whereas Gounaris et al \cite{Gounaris:1999kf} have chosen the convention where $\epsilon^{0123}=+1$.}.
\begin{equation} \label{eq:f_4Z}
e\Gamma_{ZZZ}^{\alpha\beta\mu}
=ie\frac{p_1^2-M_Z^2}{M_Z^2}\left[f_4^Z(p_1^{\alpha}g^{\mu\beta}+p_1^{\beta}g^{\mu\alpha})
+f_5^Z\epsilon^{\mu\alpha\beta\rho}\ell_\rho\right],
\end{equation}
where 
\begin{equation} \label{eq:define-ell}
\ell\equiv p_2-p_3 \equiv 2p_2-p_1
\end{equation}
with $e$ being the proton charge, the momenta ($p_1$ incoming and $p_2,p_3$ outgoing) and Lorentz indices as defined in Fig.~\ref{fig:blob-zzz}.
The dimensionless form factor $f_4^Z$ violates CP while $f_5^Z$ conserves CP.

Our aim is to determine the CP-violating contributions to the $ZZZ$ vertex, hence the contributions to $f_4^Z$. Let us here make some qualitative comments. 
Summing over $i,j,k$ (see Fig.~\ref{fig:Feynman-j2-a-app} in Appendix~A) one might think that contributions to the triangle diagram would pairwise cancel because of the factor $\epsilon_{ijk}$. 
Indeed, the scalar triangle diagrams do sum to zero, but there are non-vanishing tensor contributions, due to the momentum factors at the $ZH_iH_j$ vertices.

Three classes of Feynman diagrams give contributions to the effective CP-violating $ZZZ$ vertex, all proportional to $\Im J_2$. They are triangle diagrams with $H_i H_j H_k$ along the internal lines, as well as diagrams where one neutral Higgs boson is replaced by a neutral Goldstone $G_0$ field, or a $Z$,
\begin{equation}
f_4^Z=f_4^{Z,HHH} + f_4^{Z,HHG}  + f_4^{Z,HHZ}.
\end{equation}
These three contributions are calculated in appendix~\ref{Sec:zzz}.

\subsection{Results}
The total one-loop contribution to $f_4^Z$ for the $ZZZ$ vertex calculated in appendix~\ref{Sec:zzz} is given by a linear combination of the three-point tensor coefficient functions $C_{001}$ and $C_{1}$ (we adopt the LoopTools notation \cite{Hahn:1998yk}) of various arguments,
\begin{align}
\label{eq:f4-zzz}
	f_4^Z(p_1^2)
	&=\frac{2\alpha}{\pi\sin^3(2\thetaW)}\frac{M_Z^2}{p_1^2-M_Z^2}
	\frac{e_1e_2e_3}{v^3} \nonumber \\
	&\times\sum_{i,j,k}\epsilon_{ijk}\bigl[
	C_{001}(p_1^2,M_Z^2,M_Z^2,M_i^2,M_j^2,M_Z^2) 
	+C_{001}(p_1^2,M_Z^2,M_Z^2,M_Z^2,M_j^2,M_k^2) \nonumber \\
	& +C_{001}(p_1^2,M_Z^2,M_Z^2,M_i^2,M_Z^2,M_k^2)
	-C_{001}(p_1^2,M_Z^2,M_Z^2,M_i^2,M_j^2,M_k^2)  \nonumber \\
	& -M_Z^2C_1(p_1^2,M_Z^2,M_Z^2,M_i^2,M_Z^2,M_k^2)
	\bigr].
\end{align}
This structure was identified 20 years ago by Chang, Keung and Pal \cite{Chang:1994cs}, who studied the set of diagrams presented in appendices~\ref{Sec:zzz-hhh} and \ref{Sec:zzz-hhg}. We find numerically that our result for the sum of these diagrams is identical to their result. There are, however, also diagrams with an internal $Z$ line, arising from the $ZZH_i$ vertex which was not included in their study. These contributions are calculated in appendix~\ref{Sec:zzz-hhz}, and numerical studies show that these are actually the dominant contributions. 

\begin{figure}[htb]
	\centerline{
		\includegraphics[width=9.0cm]{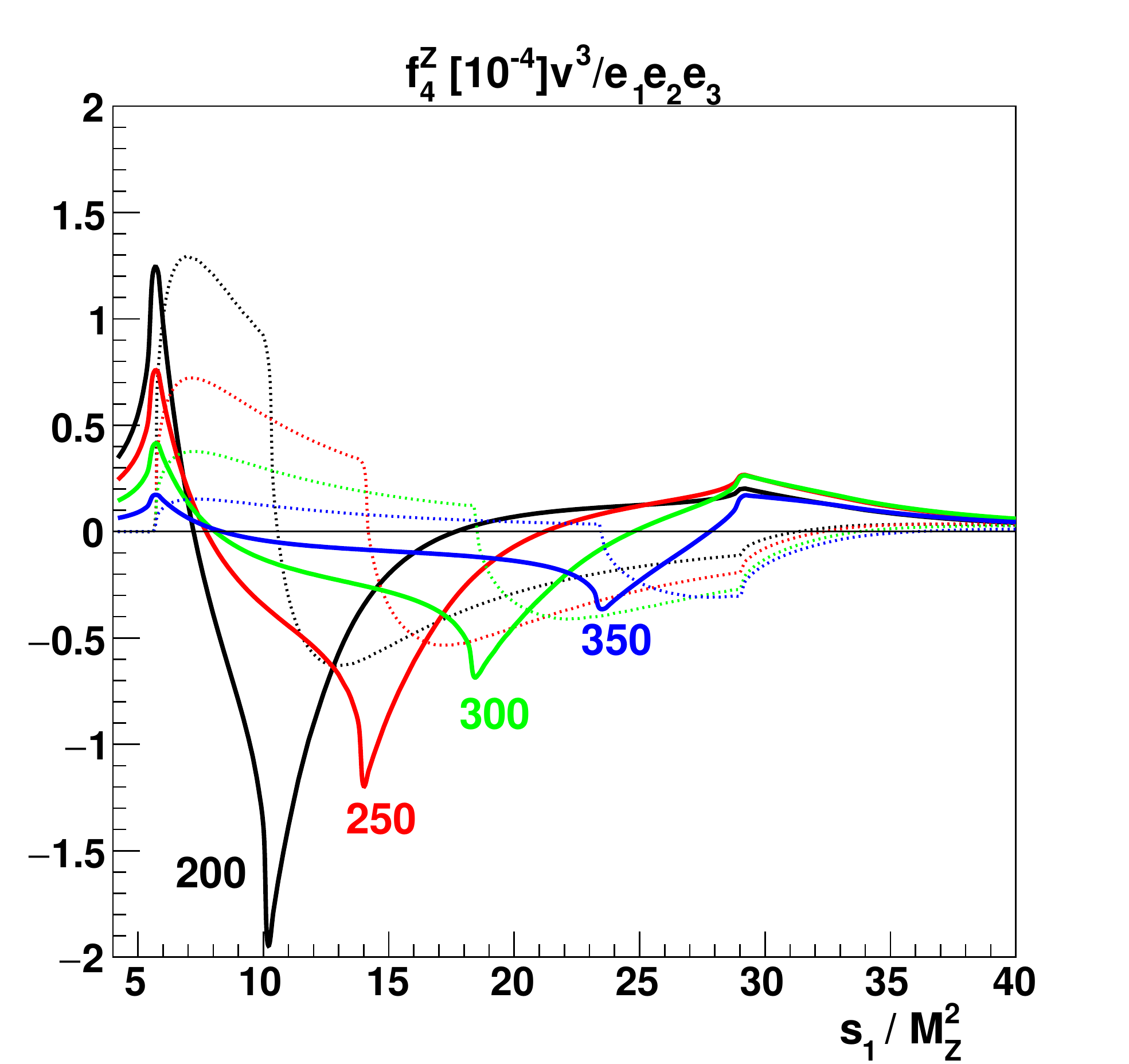}}
	\caption{Real (solid lines) and imaginary (dashed) part of the form factor $f_4^Z$ (divided by $e_1e_2e_3/v^3$) as a function of $p_1^2/M_Z^2$, for $p_2^2=p_3^2=M_Z^2$ and four values of neutral-Higgs masses $M_2$ of Eq.~(\ref{eq:higgs-masses}), as indicated (in GeV). Below threshold, $s_1=p_1^2=4M_Z^2$, the function is not defined.}
	\label{fig:c001-s1}
\end{figure}

For the neutral-Higgs masses
\begin{equation} \label{eq:higgs-masses}
	M_1=125~\text{GeV}, \quad M_2=(200, 250, 300, 350)~\text{GeV}, \quad M_3=400~\text{GeV},
\end{equation}
we show in Fig.~\ref{fig:c001-s1} the value of $f_4^Z(p_1^2)v^3/(e_1e_2e_3)$ as a function of $p_1^2/M_Z^2$. The normalization factor, $e_1e_2e_3/v^3$, is typically of ${\cal O}(0.1)$
(only small regions of the parameter space are compatible with theoretical and experimental constraints \cite{Basso:2012st,Basso:2013wna}). 
Defining $\delta$ as a measure of deviation of the $H_1VV$ coupling from its SM strength, $e_1=v(1-\delta)$, and using $e_2^2+e_3^2=v^2-e_1^2$, one can easily find~\cite{Grzadkowski:2014ada}
that for small $\delta$, $(e_1e_2e_3)/v^3 < \delta$, so it is suppressed by the $H_1VV$ coupling approaching the SM limit.

The form factor $f_4^Z$ has been constrained by experiments at LEP, Fermilab and the LHC.
Recently, CMS \cite{Khachatryan:2015pba} has presented an impressive bound on $f_4^Z$ (assumed real): $-0.0022<f_4^Z<0.0026$. This result is obtained in the $2\ell2\nu$ channel from the 7 and 8~TeV data sets.
It is still two orders of magnitude above what is generated in the 2HDM by a non-zero $\Im J_2$.

\section{The \boldmath{$ZW^+W^-$} vertex}
\label{Sec:ZWW}
\setcounter{equation}{0}
\begin{figure}[htb]
	\vspace*{-2.0cm}
	\centerline{
		\includegraphics[width=15.0cm]{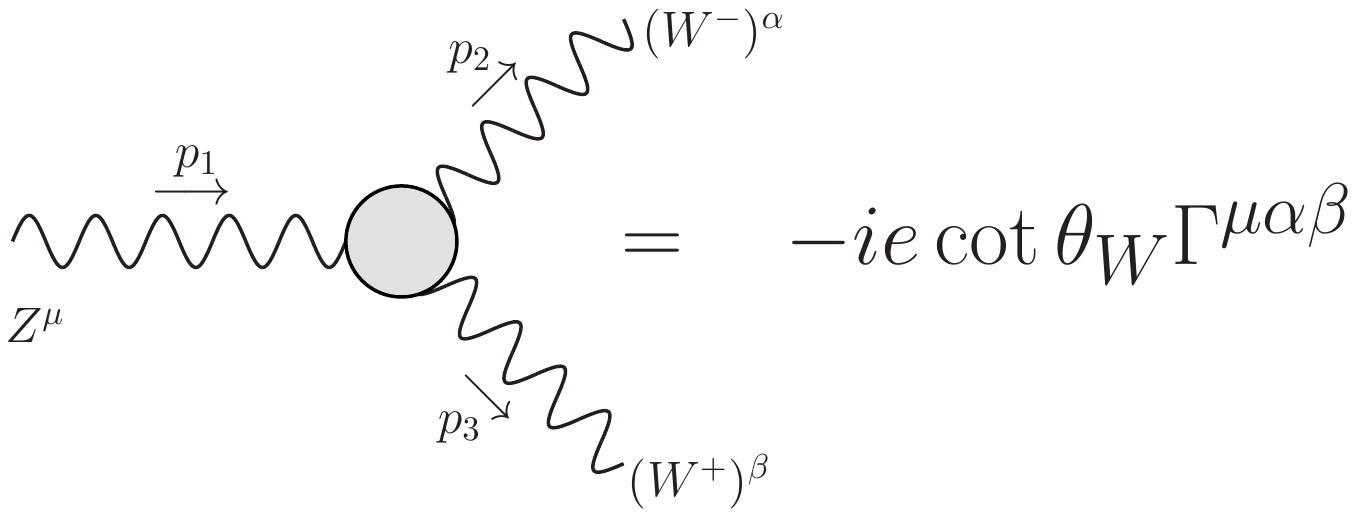}}
	\vspace*{-16.0cm}
	\caption{The general $ZWW$ vertex.}
	\label{fig:blob-zww}
	
\end{figure}

Contrary to the $ZZZ$ vertex, the $ZWW$ vertex is present at the tree level, with a well-known, CP-conserving structure:
\begin{subequations}
\begin{align}
ig_{ZWW}\Gamma^{\alpha\beta\mu}_\text{tree}
&=-ig\cos\theta_W[g^{\alpha\beta}(p_2-p_3)^\mu + g^{\beta\mu}(p_1+p_3)^\alpha 
- g^{\mu\alpha}(p_1+p_2)^\beta] \\
&=-ig\cos\theta_W[g^{\alpha\beta}\ell^\mu + g^{\beta\mu}(-\half\ell+\threehalf p_1)^\alpha
- g^{\mu\alpha}(\half\ell +\threehalf p_1)^\beta],
\end{align}
\end{subequations}
where $g_{ZWW}=-e\cot\theta_W$, $p_1$ is incoming while $p_2$ and $p_3$ are outgoing, and in the second line, we make use of $\ell=p_2-p_3$. 

Triangle diagrams discussed in appendix~\ref{Sec:zww} contribute to the CP-violating $ZW^+W^-$ vertex. In fact, they give a contribution proportional to the invariant $\Im J_2$, which is one measure of CP violation in the Two-Higgs-Doublet model \cite{Gunion:2005ja} (referred to as $J_1$ in earlier work by Lavoura, Silva and Botella \cite{Lavoura:1994fv,Botella:1994cs}).

\subsection{Lorentz structure}
Phenomenological discussions \cite{Hagiwara:1986vm} of the $ZWW$ vertex have presented its most general Lorentz structure. We let $Z$ be off-shell while both $W^\pm$ are on-shell, again assuming that $Z$ couples to a pair of leptons $e^+e^-$ so that we may neglect terms proportional to the lepton mass. Then according to \cite{Hagiwara:1986vm} the structure reads
\begin{align} \label{eq:hagiwara2}
\Gamma_{ZWW}^{\alpha\beta\mu}
&=f_1^Z\ell^{\mu}g^{\alpha\beta}
-\frac{f_2^Z}{M_W^2}\ell^{\mu}p_1^{\alpha}p_1^{\beta}
+f_3^Z(p_1^{\alpha}g^{\mu\beta}-p_1^{\beta}g^{\mu\alpha}) \nonumber \\
&+if_4^Z(p_1^{\alpha}g^{\mu\beta}+p_1^{\beta}g^{\mu\alpha})
+if_5^Z\epsilon^{\mu\alpha\beta\rho}\ell_\rho \nonumber \\
&-f_6^Z\epsilon^{\mu\alpha\beta\rho}p_{1\rho}
-\frac{f_7^Z}{M_W^2}\ell^{\mu}\epsilon^{\alpha\beta\rho\sigma}p_{1\rho}\ell_\sigma.
\end{align}
The tree-level vertex contributes to $f_1$ and $f_3$: 
\begin{equation}
f_1^\text{tree}=1, \quad
f_3^\text{tree}=2.
\end{equation}
The dimensionless form factors $f_4^Z$, $f_6^Z$ and $f_7^Z$ violate CP while the others conserve CP. Recent LHC experiments \cite{Chatrchyan:2013yaa,Aad:2014mda,Khachatryan:2015sga} have constrained the CP-conserving anomalous couplings, but not the CP-violating $f_4^Z$.

Our aim is to determine the CP violating contributions to the $ZWW$ vertex, hence the contributions to $f_4^Z$.

\subsection{Results}
The total one-loop contribution to $f_4^Z$ for the $ZWW$ vertex calculated in appendix~\ref{Sec:zww} is given by a linear combination of the three-point tensor coefficient functions $C_{001}$ of various arguments,
\begin{align} \label{eq:f_4W}
	f_4^Z(p_1^2)
	=\frac{-\alpha}{\pi\sin^2(2\thetaW)}
	\frac{e_1e_2e_3}{v^3} 
	\sum_{i,j,k}\epsilon_{ijk}\bigl[
	&C_{001}(p_1^2,M_W^2,M_W^2,M_i^2,M_j^2,M_W^2) \nonumber \\
	-&C_{001}(p_1^2,M_W^2,M_W^2,M_i^2,M_j^2,M_{H^\pm}^2) 
	\bigr].
\end{align}
This quantity was also studied by He, Ma and McKellar \cite{He:1992qh}. Assuming that they have used the $(-i\epsilon)$ prescription in their Eq.~(5), we find numerical agreement apart from an overall sign. Furthermore, the result for the imaginary part given in their Eq.~(6) is twice as large as the one in Eq.~(5).

\begin{figure}[htb]
	\centerline{
		\includegraphics[width=9.0cm]{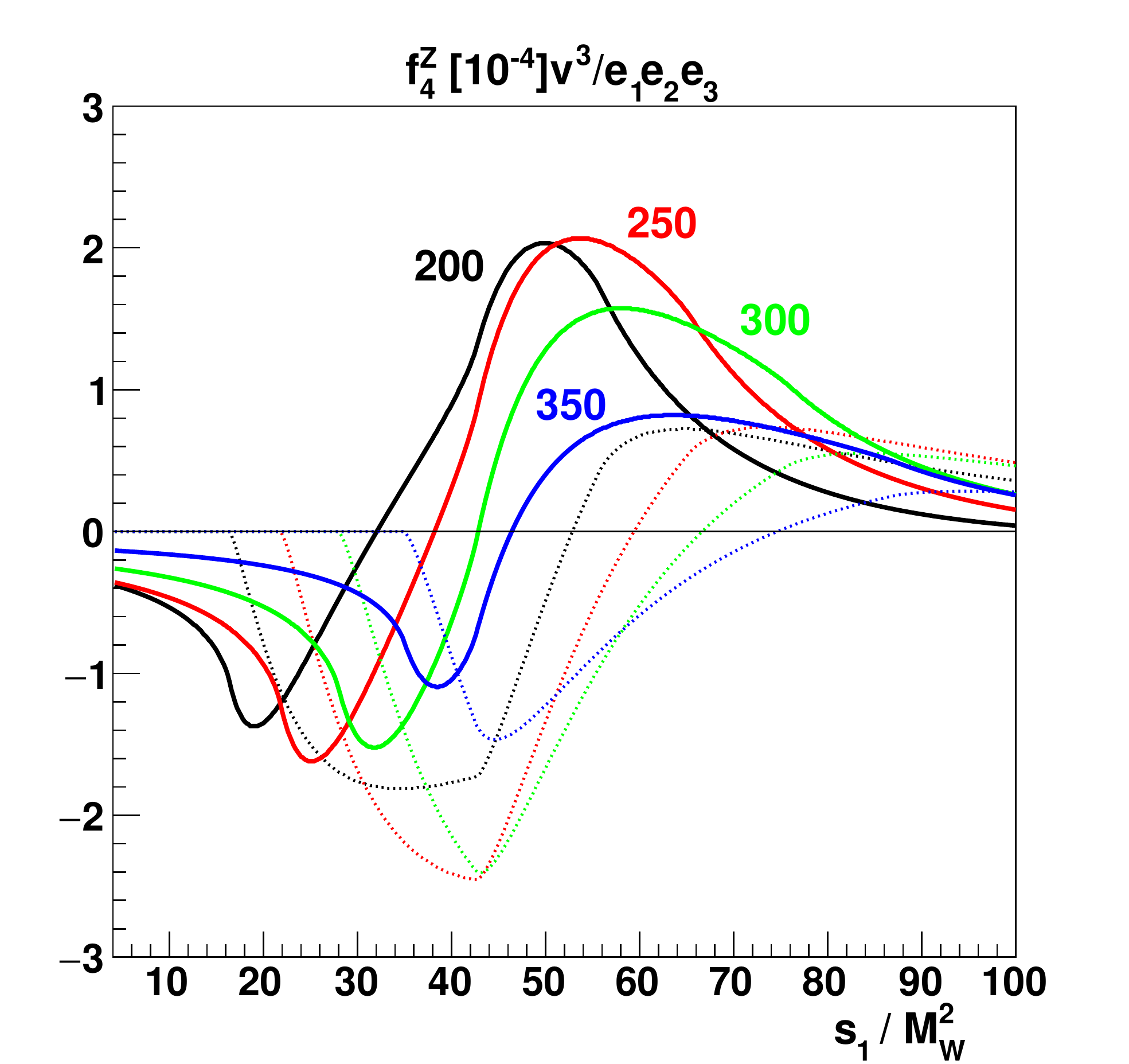}}
	\caption{Real (solid lines) and imaginary (dashed) part of the form factor $f_4^Z$ (divided by $e_1e_2e_3/v^3$) as a function of $s_1/M_W^2$, for $p_2^2=p_3^2=M_W^2$ and four values of neutral-Higgs $M_2$ masses of Eq.~(\ref{eq:higgs-masses}), as indicated (in GeV).}
	\label{fig:c001-s1-zww}
\end{figure}

For the neutral-Higgs masses given by equation~(\ref{eq:higgs-masses}),
we show in Fig.~\ref{fig:c001-s1-zww} the value of $f_4^Z(p_1^2)v^3/(e_1e_2e_3)$ as a function of $s_1/M_W^2$. 

\section{Asymmetries}
\label{Sec:asym}
\setcounter{equation}{0}
We are going to discuss the possibility of testing CP violation at future $e^+e^-$ colliders \cite{Djouadi:2007ik,Lebrun:2012hj}.
It is assumed that polarizations of the final-state vector bosons could be determined experimentally\footnote{Investigating angular distributions of the vector boson decay products one can indeed measure their polarizations.}. We adopt 
CP-sensitive observables defined for $W^+W^-$ and $ZZ$ in \cite{Gounaris:1991ce,Chang:1993vv}, and \cite{Chang:1994cs}, respectively. Below, we present some predictions for those and other asymmetries for the 2HDM.

\subsection{\boldmath{$e^+e^-\to ZZ$}}
\label{Sec:asymzz}
Helicities of the $ZZ$ (and $W^+W^-$) pairs can be measured statistically by studying decay products of the final vector bosons.
Therefore, we will define a number of differential asymmetries assuming that both the momenta and helicities of the $ZZ$ pair could be determined.
Since our goal is to measure the CP-violating form factor $f_4^Z$, our asymmetries will (to leading order) be proportional to $f_4^Z$.
Let us first start by considering
\bea
A_1^{ZZ}\equiv\frac{\sigma_{+,0}-\sigma_{0,-}}{\sigma_{+,0}+\sigma_{0,-}},\\
A_2^{ZZ}\equiv\frac{\sigma_{0,+}-\sigma_{-,0}}{\sigma_{0,+}+\sigma_{-,0}},
\eea
where $\sigma_{\lambda,\bar{\lambda}}$ are unpolarized-beam cross sections for the production of $ZZ$ with helicities $\lambda$ and $\bar{\lambda}$, respectively. 
The cross sections can be expressed through the helicity amplitudes
for $e^+(\sigma)e^-(\bar\sigma)\to Z(\lambda)Z(\bar\lambda)$ as follows 
 \beq
 \sigma_{\lambda,\bar\lambda}=\sum_{\sigma,\bar\sigma} 
 \mcal_{\sigma,\bar\sigma;\lambda\bar\lambda}(\Theta)\mcal^\star_{\sigma,\bar\sigma;\lambda\bar\lambda}(\Theta),
 \label{xsec}
 \eeq
where $\sigma$ and $\bar\sigma$ are the helicities of $e^-$ and $e^+$, respectively.
Expressions for these cross sections can readily be written out using the results from  Chang, Keung and Pal \cite{Chang:1994cs}. Letting $\Theta$ be the angle between the $e^{-}$ beam direction and the $Z$ whose helicity is given by the first index $\lambda$, and defining $\gamma=\sqrt{s_1}/(2M_Z)$ and $\beta^2=1-\gamma^{-2}$, we find to lowest order in $f_4^Z$
\begin{equation}
A_1^{ZZ}=
-4 \beta \gamma^4 \left[(1+\beta ^2)^2-(2 \beta \cos \Theta)^2 \right]
{\cal F}_1(\beta,\Theta)\,\Im f_4^Z,
\label{a1zz}
\end{equation}
with ${\cal F}_1(\beta,\Theta)$ given in appendix~D.

\begin{figure}[htb]
	\centerline{
		\includegraphics[width=8.0cm]{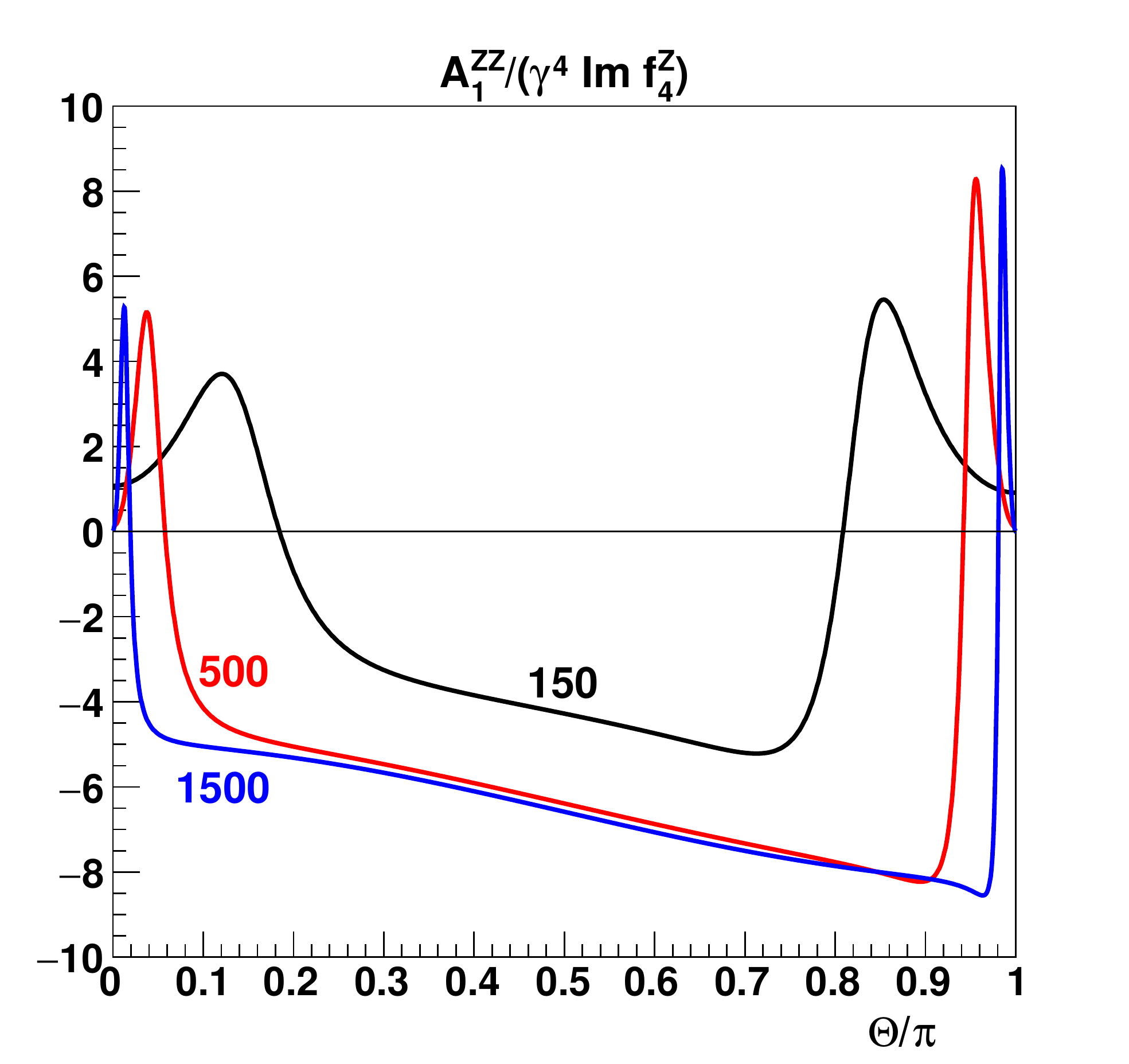}
		\includegraphics[width=8.0cm]{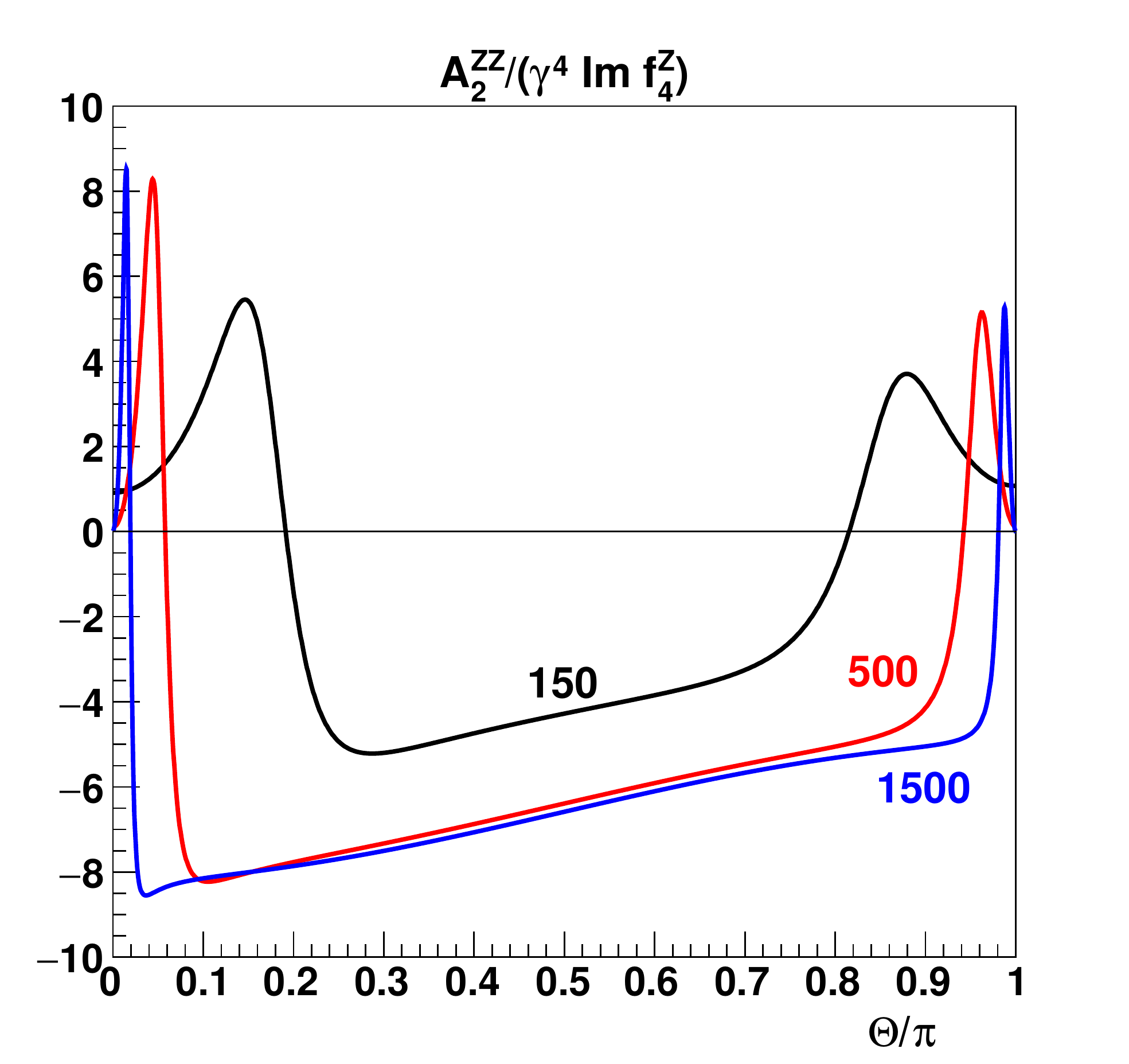}}
	\caption{The asymmetries $A_1^{ZZ}(\Theta)$ of Eq.~(\ref{a1zz}) and  $A_2^{ZZ}(\Theta)$ of Eq.~(\ref{a2zz})(both divided by $\gamma^4\,\Im f_4^Z$) as functions of $\Theta$ for three beam energies $E$ as indicated (in GeV).}
	\label{fig:A1-A2_zz}
\end{figure}

In the low-energy limit ($\beta\to 0$) this simplifies to
\bea
A_1^{ZZ}&=&\frac{-4 \beta \left[\xi _1-3 \xi _1 \cos^2\Theta +2 \left(\xi _1-\xi _2\right) \cos^3\Theta \right]\Im f_4^Z}
{(\xi _3+\xi _4)+2 \xi _3 \cos\Theta-3 \left(\xi _3+\xi _4\right) \cos^2\Theta -4 \xi _3 \cos^3\Theta +4 \left(\xi _3+\xi _4\right) \cos^4\Theta},\nonumber\\
\eea
where the $\xi_i$ are given in appendix~D.
Furthermore, we find
\begin{equation}
A_2^{ZZ}=A_1^{ZZ}\left(\cos \Theta\to-\cos \Theta\right)\label{a2zz}.
\end{equation}
These asymmetries are both shown in Fig.~\ref{fig:A1-A2_zz}.
The sharp peaks near the forward and backward directions are due to an interplay of three factors: (1) the near-divergence of the $t$-channel propagator, (2) the factor $[\Delta\sigma\Delta\lambda(1+\beta^2)-2\cos\Theta]$ of the amplitude (see Eq.~(5) in ref.~\cite{Chang:1994cs}) and (3) the Wigner functions proportional to $1\pm\cos\Theta$.

\begin{figure}[htb]
	\centerline{
		\includegraphics[width=8.0cm]{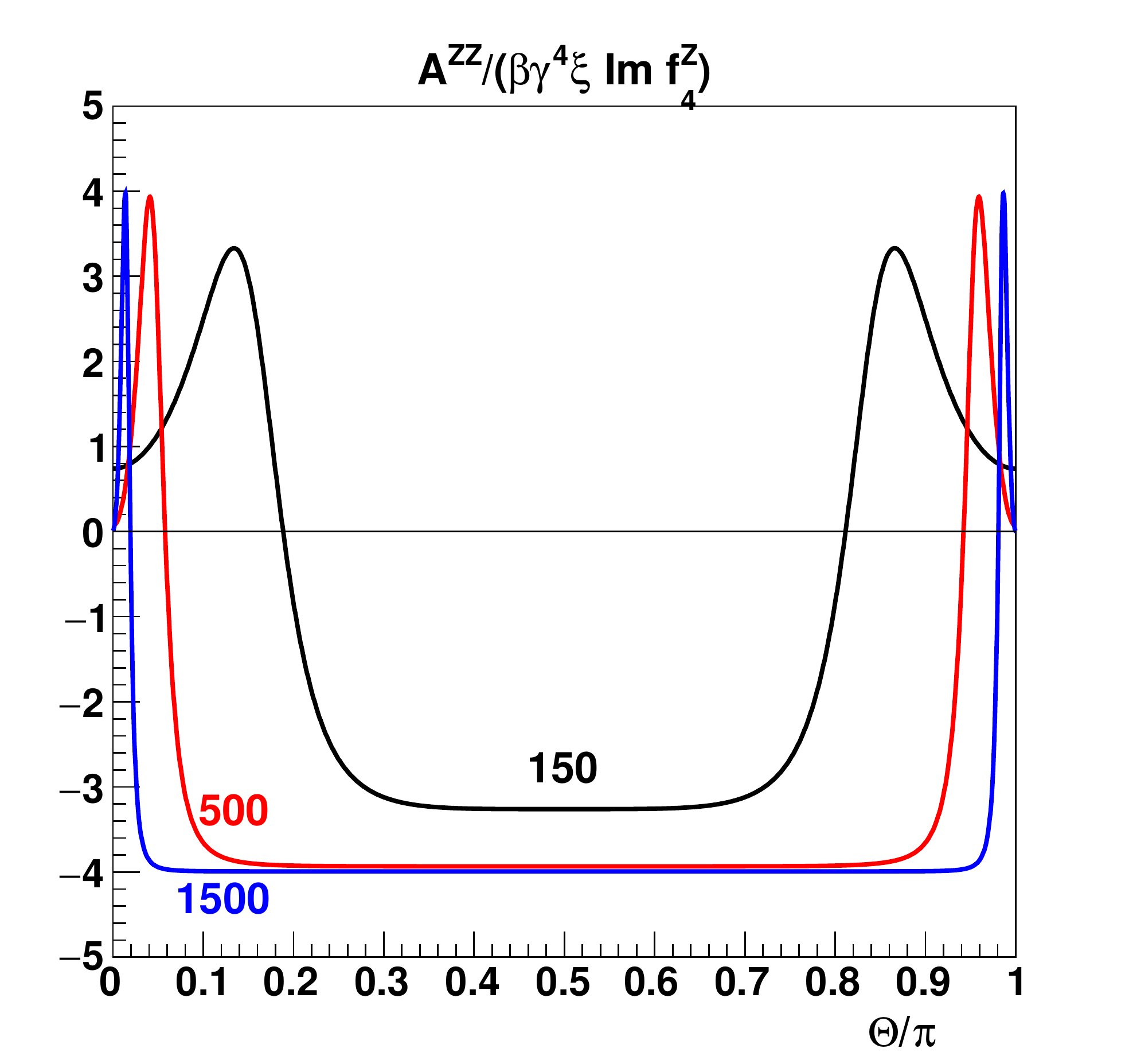}
		\includegraphics[width=8.0cm]{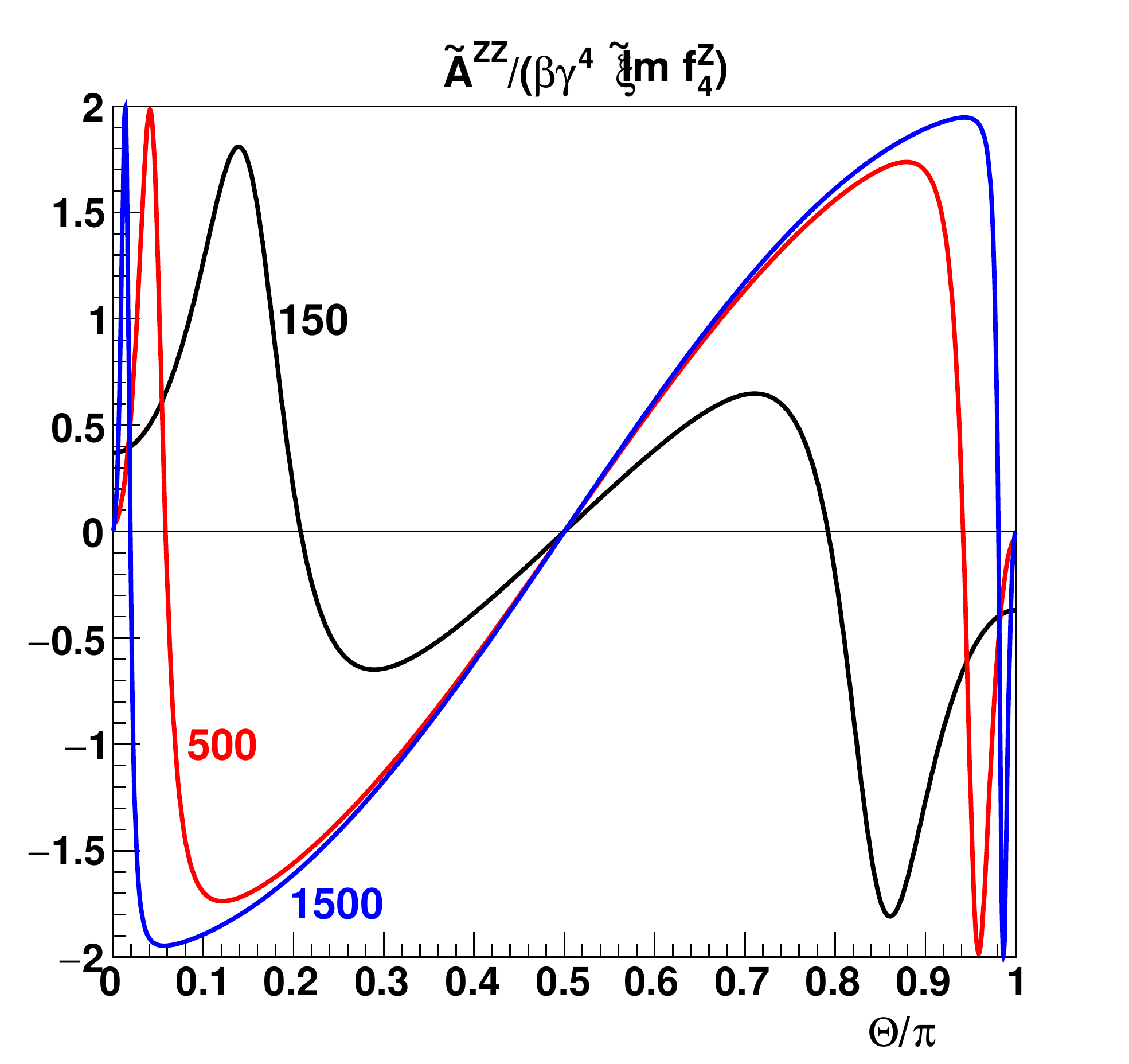}}
	\caption{The asymmetries $A^{ZZ}(\Theta)$ and $\tilde{A}^{ZZ}(\Theta)$ of Eqs.~(\ref{azz}) and (\ref{atildezz}) (divided by $\beta\gamma^4\xi\,\Im f_4^Z$ and $\beta\gamma^4\tilde \xi\,\Im f_4^Z$) and  as functions of $\Theta$ for three beam energies $E$ as indicated (in GeV).}
	\label{fig:A_zz}
\end{figure}

Introducing the abbreviations
\begin{eqnarray}
	\xi&=&\frac{2\sin\theta_W\cos\theta_W(1-6\sin^2\theta_W+12\sin^4\theta_W)}{1-8\sin^2\theta_W+24\sin^4\theta_W-32\sin^6\theta_W+32\sin^8\theta_W}
	\simeq 1.65,\\
	\tilde{\xi}&=&\frac{-4 \sin \theta _W \cos \theta _W\left(1-6 \sin ^2\theta _W+12 \sin ^4\theta _W-16 \sin ^6\theta _W\right) }
	{1-8 \sin ^2\theta _W+24 \sin ^4\theta _W-32 \sin ^6\theta _W+32 \sin ^8\theta _W}
		\simeq -0.78,
\end{eqnarray}
the following asymmetries can be defined and calculated to leading order in $f_4^Z$:
\begin{eqnarray} 
A^{ZZ}&\equiv&\frac{\sigma_{+,0}+\sigma_{0,+}-\sigma_{0,-}-\sigma_{-,0}}{\sigma_{+,0}+\sigma_{0,+}+\sigma_{0,-}+\sigma_{-,0}} \nonumber \\
&=&\frac{-2\beta\gamma^4[(1+\beta^2)^2-(2\beta\cos\Theta)^2][1+\beta^2-(3-\beta^2)\cos^2\Theta]\xi\,\Im f_4^Z}
{(1+\beta^2)^2-(3+6\beta^2-\beta^4)\cos^2\Theta+4\cos^4\Theta},\\
\label{azz} 
\tilde{A}^{ZZ}&\equiv&\frac{\sigma_{+,0}-\sigma_{0,+}-\sigma_{0,-}+\sigma_{-,0}}{\sigma_{+,0}+\sigma_{0,+}+\sigma_{0,-}+\sigma_{-,0}} \nonumber \\
&=&\frac{-2\beta  \gamma ^4  \cos \Theta[(1+\beta^2)^2-(2\beta\cos\Theta)^2] \left(\beta ^2- \cos ^2\Theta \right)\tilde{\xi}\Im f_4^Z }
{\left(1+\beta ^2\right)^2-\left(3+6 \beta ^2-\beta ^4\right) \cos ^2\Theta +4 \cos ^4\Theta }.
\label{atildezz} 
\end{eqnarray}
The asymmetries $A^{ZZ}$ and $\tilde A^{ZZ}$ are both shown in Fig.~\ref{fig:A_zz} for three values of the energy.
Since the former is defined symmetrically with respect to the two $Z$ bosons, the expression is forward-backward symmetric.
At high energies and intermediate angles, it is well approximated by $A^{ZZ}\simeq -4\gamma^4\xi\, \Im f_4^Z$.

In the low-energy limit, these become
\begin{eqnarray}
A^{ZZ}&\to& \frac{-2\beta(1-3\cos^2\Theta)\xi\,\Im f_4^Z}
{1-3\cos^2\Theta+4\cos^4\Theta},\\
\tilde{A}^{ZZ}&\to& \frac{2\beta\cos^3\Theta\tilde{\xi}\,\Im f_4^Z}
{1-3\cos^2\Theta+4\cos^4\Theta}.
\end{eqnarray}

\begin{figure}[htb]
	\centerline{
		\includegraphics[width=9.0cm]{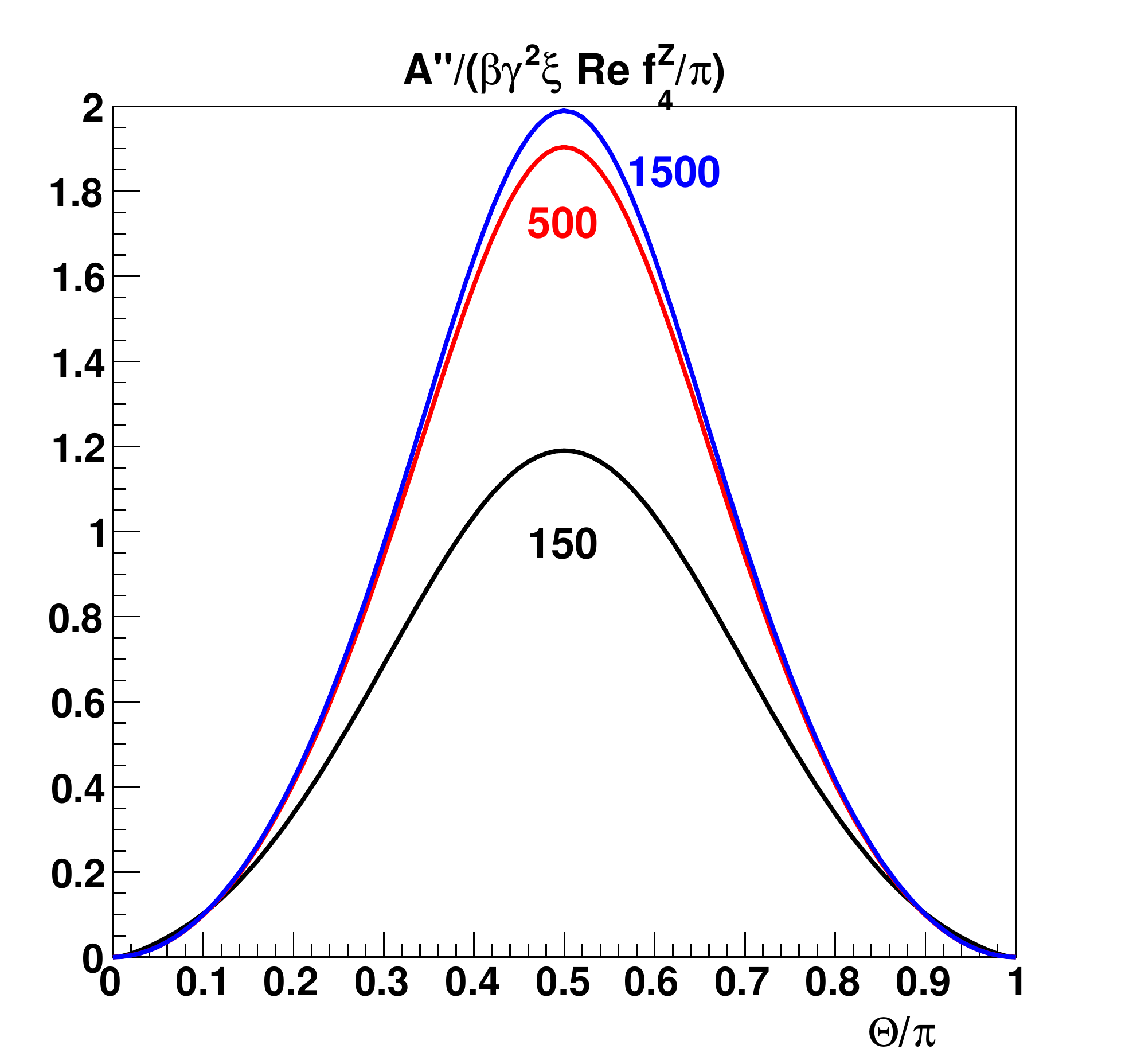}}
	\caption{The asymmetry ${\cal A}^{\prime\prime}(\Theta)$ of Eq.~(\ref{Eq:A_bis}) (divided by $\beta\gamma^2\xi\,\Re f_4^Z/\pi$) as a function of $\Theta$ for three beam energies $E$ as indicated (in GeV).}
	\label{fig:A_bis}
\end{figure}

Other possibilities of testing CP violation in $e^+e^-\to ZZ$ have been investigated by Chang, Keung and Pal \cite{Chang:1994cs}, who note that the angular distribution of $\ell^-$ from a $Z$ decay is determined by the spin-density matrix of the $Z$ (see Eq.~(10) of Ref.~\cite{Chang:1994cs}):
\begin{equation}
\rho(\Theta)_{\lambda_1\lambda_2}={\cal N}^{-1}(\Theta)
\sum_{\sigma,\bar\sigma,\bar\lambda}{\cal M}_{\sigma,\bar\sigma,\lambda_1,\bar\lambda}(\Theta)
{\cal M}^\ast_{\sigma,\bar\sigma,\lambda_2,\bar\lambda}(\Theta).
\end{equation}
where again, $\sigma$ and $\bar\sigma$ are helicities of $e^-$ and $e^+$, respectively, and the $\lambda$ and $\bar\lambda$ refer to the two $Z$ helicities.
They advocate a certain difference of cross sections, integrated over azimuthal quadrants of the final-state leptons, which is not suppressed by the approximate C-symmetry.
They have thus defined such a ``folded'' asymmetry ${\cal A}^{\prime\prime}(\Theta)$ in their Eq.~(15), and shown that it equals
\begin{equation}
	{\cal A}^{\prime\prime}(\Theta)
	=-\frac{1}{\pi}
	\left[\Im\rho(\Theta)_{+,-}-\Im\rho(\pi-\Theta)_{-,+}\right].
\end{equation}
To lowest order in $f_4^Z$, this quantity is proportional to $\Re f_4^Z$:
\begin{equation} \label{Eq:A_bis}
	{\cal A}^{\prime\prime}(\Theta)
	=\frac{\beta(1+\beta^2)\gamma^2
		[(1+\beta^2)^2-(2\beta\cos\Theta)^2]\sin^2\Theta\, \xi\, \Re f_4^Z}
	{\pi[2+3\beta^2-\beta^6
		-\beta^2(9-10\beta^2+\beta^4)\cos^2\Theta
		-4\beta^4\cos^4\Theta]}.
\end{equation}
This asymmetry is shown in Fig.~\ref{fig:A_bis} for three values of the energy.
Superficially, it looks like this asymmetry might be unbounded at high energies. This is not the case, since at high energies (see Appendix~C) $f_4^Z$ falls off like $(1/\gamma^6)\log\gamma$.

In the low-energy limit ($\beta\to0$), it simplifies:
\begin{equation}
	{\cal A}^{\prime\prime}(\Theta) \to \frac{\beta\sin^2\Theta\,\xi \,\Re f_4^Z}{2\pi}.
\end{equation}

\subsection{\boldmath{$e^+e^-\to W^+W^-$}}
\label{Sec:asymww}
Let us follow the same approach as for $e^+e^-\to ZZ$ in the $e^+e^-\to W^+W^-$ case by forming the asymmetries \cite{Chang:1993vv}:
\bea
A_1^{WW}\equiv\frac{\sigma_{+,0}-\sigma_{0,-}}{\sigma_{+,0}+\sigma_{0,-}},
\label{a1} \\
A_2^{WW}\equiv\frac{\sigma_{0,+}-\sigma_{-,0}}{\sigma_{0,+}+\sigma_{-,0}},
\label{a2}
\eea
where $\sigma_{\lambda,\bar{\lambda}}$ are unpolarized-beam cross sections for the production of $W^-$ and $W^+$ with helicities $\lambda$ and $\bar{\lambda}$, respectively. 
The cross sections can be expressed through the helicity amplitudes
 for $e^+(\sigma)e^-(\bar\sigma)\to W^-(\lambda)W^+(\bar\lambda)$ like in Eq.~(\ref{xsec}), where $\sigma$ and $\bar\sigma$ are the helicities of $e^-$ and $e^+$, respectively.
The amplitudes $\mcal_{\sigma,\bar\sigma;\lambda\bar\lambda}(\Theta)$
 were first calculated in \cite{Hagiwara:1986vm}.
 Here, $\Theta$ is the angle between the $e^-$ and the $W^-$ momenta.

 Following the notation of \cite{Chang:1993vv}, we find for the case of polarized initial beams ($\sigma,\bar\sigma$), and to lowest order in $f_4^Z$:
 \begin{subequations}
 	\begin{alignat}{2}
 	&(\sigma,\bar\sigma)=(+-):&\quad A_1^{WW}&=\frac{s_1}{M_Z^2}\Im f_4^Z, \\
 	&(\sigma,\bar\sigma)=(-+):&\quad A_1^{WW}&=\frac{-\beta^2(1-2\sin^2\thetaW)s_1}
	{\beta^2(2\sin^2\thetaW M_Z^2-s_1)
 		+(s_1-M_Z^2)Y}\Im f_4^Z,
 	\end{alignat}
 \end{subequations}
 where
 \beq
 Y \equiv 1 - \frac{(1+\beta)}{\gamma^2(1+\beta^2-2\beta \cos\Theta)}.
 \eeq
 with $\gamma=\sqrt{s_1}/(2M_W)$ and $\beta^2=1-\gamma^{-2}$.
 
 For the unpolarized case, we find (still to lowest order in $f_4^Z$):
 \begin{equation}
 A_1^{WW}=
 \frac{N_1^{(a)}(1-\cos\Theta)^2+N_1^{(b)}(1+\cos^2\Theta)}
 {D_1^{(a)}(1-\cos\Theta)^2+D_1^{(b)}(1+\cos^2\Theta)}\,
 \beta s_1\,\Im f_4^Z
 \end{equation}
 with  the following abbreviations:
 \begin{subequations} \label{Eq:asymmetry-details}
 	\begin{align}
 	N_1^{(a)}&=(1+\beta^2-2\beta\cos\Theta)
 	\{X_1 -2\sin^2\thetaW[(1-\beta^2)(1-\beta+2\cos\Theta)s_1 \nonumber \\
 	& -(1-3\beta-\beta^2+2\beta^2\cos\Theta-\beta^3+2\cos\Theta)M_Z^2]
 	\},  \\
 	N_1^{(b)}&=8\sin^4\thetaW\beta(1+\beta^2-2\beta\cos\Theta)^2M_Z^2, \\
 	D_1^{(a)}&=X_1^2-4\sin^2\thetaW\beta(1+\beta^2-2\beta\cos\Theta) X_1M_Z^2, \\
 	D_1^{(b)}&=8\sin^4\thetaW\beta^2(1+\beta^2-2\beta\cos\Theta)^2M_Z^4, \\
 	X_1&=(1-\beta^2)(1-\beta+2\cos\Theta)s_1-(1-2\beta-\beta^2+2\cos\Theta)M_Z^2.
 	\end{align}
 \end{subequations}
 
 \begin{figure}[htb]
 	\centerline{
 		\includegraphics[width=8.0cm]{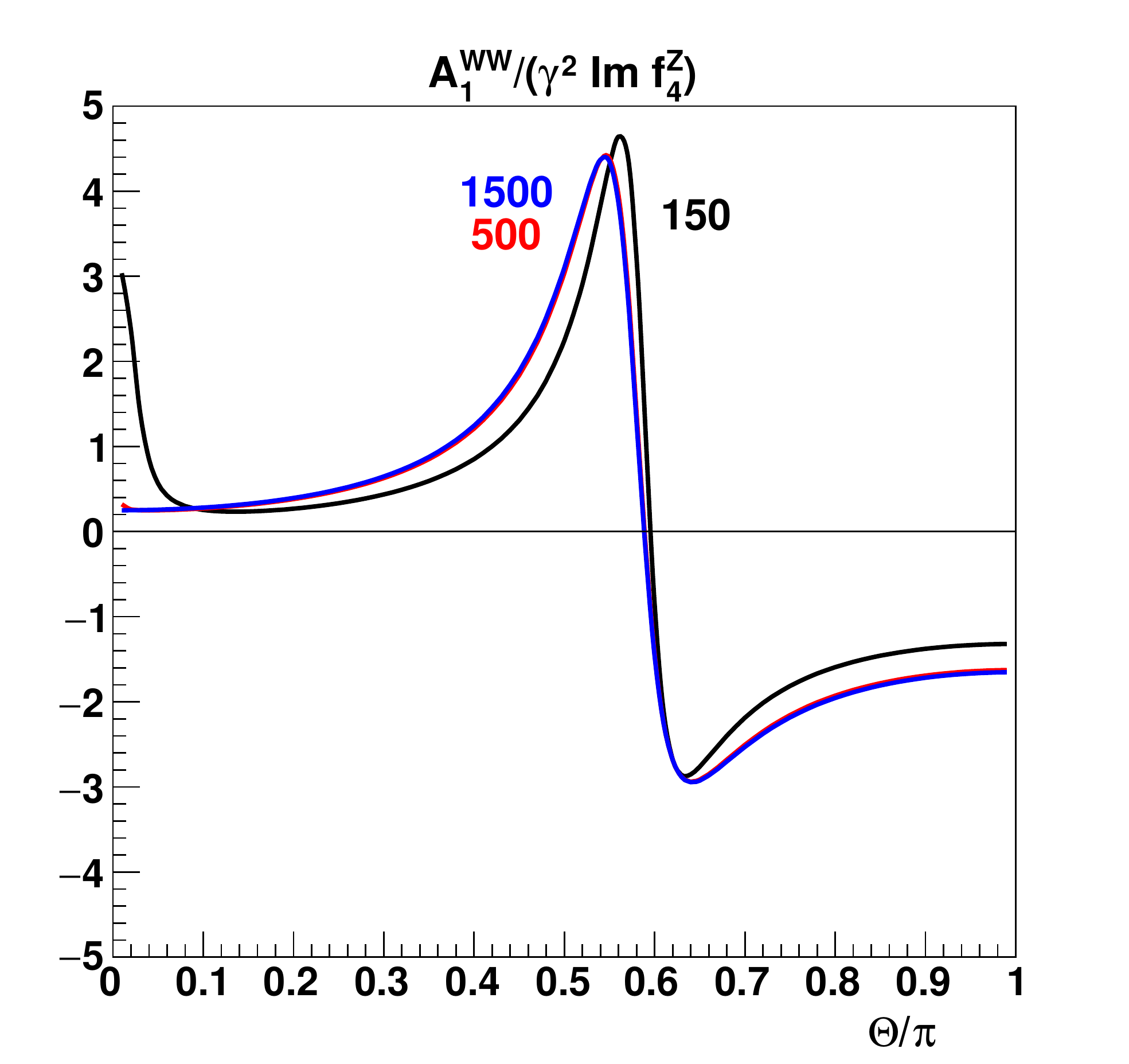}
		\includegraphics[width=8.0cm]{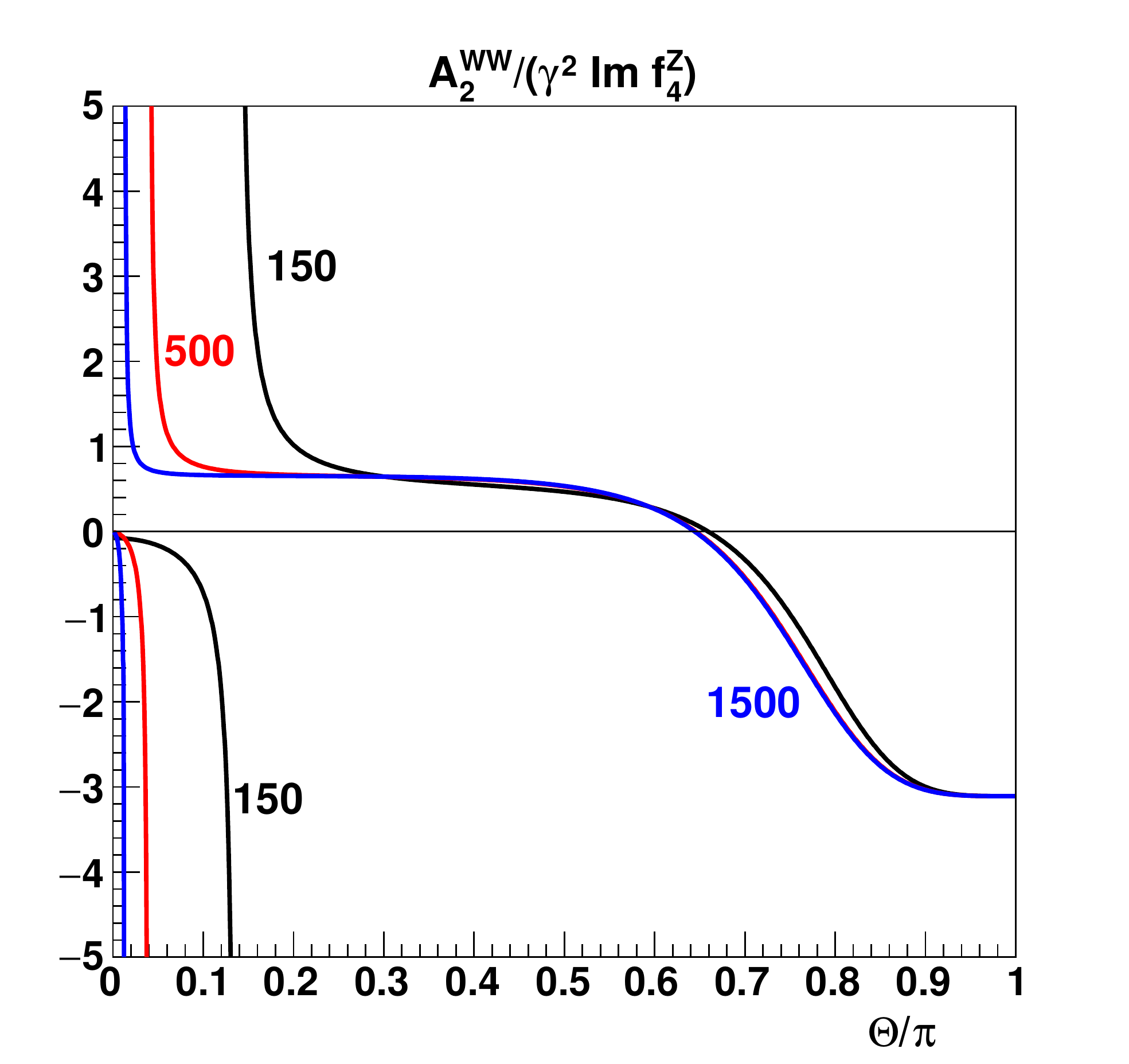}}
 	\caption{The asymmetries $A_1^{WW}$ and $A_2^{WW}$ vs $\Theta$ for three values of the beam (or $W$) energy $E$, 150~GeV, 500~GeV and 1500~GeV, as indicated.}
 	\label{fig:a_1-a_2_ww}
 \end{figure}
  
 In the low-energy limit ($\beta\to0$), this simplifies:
 \begin{equation}
 A_1^{WW} \to\begin{cases} \frac{4M_W^2}{M_Z^2}\frac{2M_W^2-M_Z^2}{(4M_W^2-M_Z^2)(1+2\cos\Theta)}\beta\,\Im f_4^Z,\hspace*{0.53cm} \beta\lsim |1+2\cos\Theta|, \quad \beta\ll1,\\
 -\frac{2 M_W^2 \left(16 M_W^4-5 M_W^2 M_Z^2-2 M_Z^4\right)}{M_Z^2 \left(10 M_W^4-2 M_W^2 M_Z^2+M_Z^4\right)}\,\Im f_4^Z,\hspace*{0.25cm} |1+2\cos\Theta|\lsim\beta\ll1,
 \end{cases}
 \end{equation}
 where we have also substituted the tree-level relation $\sin^2\thetaW=1-M_W^2/M_Z^2$.
  
 Furthermore, we find
 \bea
 A_2^{WW}&=&-A_1^{WW}\left(\cos \Theta\to-\cos \Theta;\beta\to -\beta\right).
 \eea

 We display these asymmetries $A_1^{WW}$ and $A_2^{WW}$ in Fig.~\ref{fig:a_1-a_2_ww}. An overall factor $\gamma^2\,\Im f_4^Z$ is factored out, and hence for $A_1^{WW}$, the graphs for 500~GeV and 1500~GeV are practically indistinguishable. The main structure is due to the first term in the numerator of Eq.~(\ref{a1}) passing through zero close to a minimum of the denominator.

We may also combine these two asymmetries into one, either by addition or subtraction. Again calculating to lowest order in $f_4^Z$: 
\begin{align} 
	A^{WW}&\equiv\frac{\sigma_{+,0}+\sigma_{0,+}-\sigma_{0,-}-\sigma_{-,0}}
	{\sigma_{+,0}+\sigma_{0,+}+\sigma_{0,-}+\sigma_{-,0}} \nonumber \\
	&=\beta\left(1+\beta^2-2 \beta  \cos \Theta\right) {\cal F}^{WW}  \,\Im f_4^Z, \label{aww}  \\
	\tilde{A}^{WW}&\equiv\frac{\sigma_{+,0}-\sigma_{0,+}+\sigma_{0,-}-\sigma_{-,0}}
	{\sigma_{+,0}+\sigma_{0,+}+\sigma_{0,-}+\sigma_{-,0}} \nonumber \\
	&=\beta\left(1+\beta^2-2 \beta  \cos \Theta\right) {\cal \tilde F}^{WW}  \,\Im f_4^Z,
	\label{awwtilde} 
\end{align}
where the functions ${\cal F}^{WW}$ and ${\cal \tilde F}^{WW}$, given in appendix~D, can be expressed as ratios of polynomials in $\cos\Theta$.

\begin{figure}[htb]
	\centerline{
		\includegraphics[width=8.0cm]{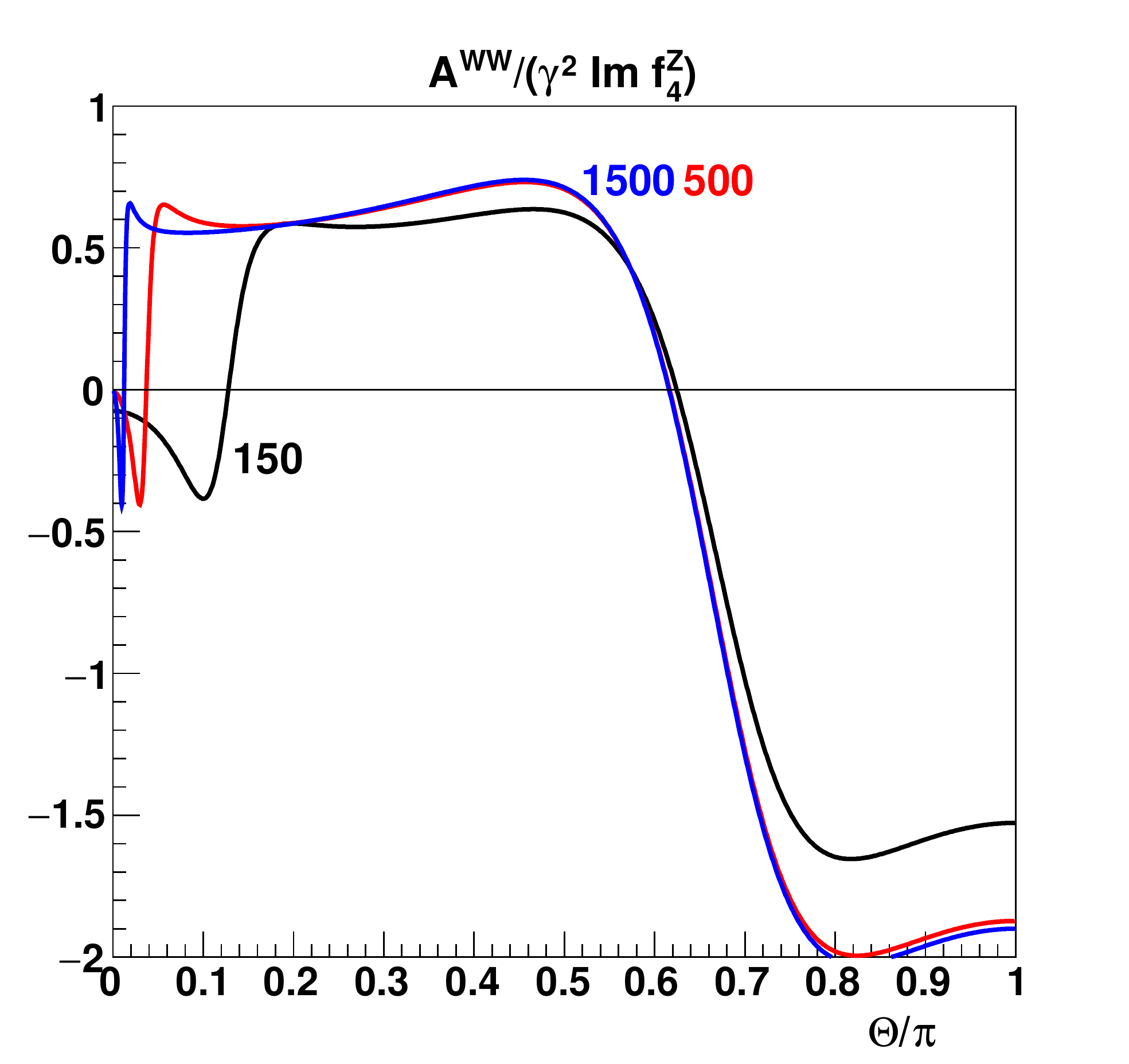}
		\includegraphics[width=8.0cm]{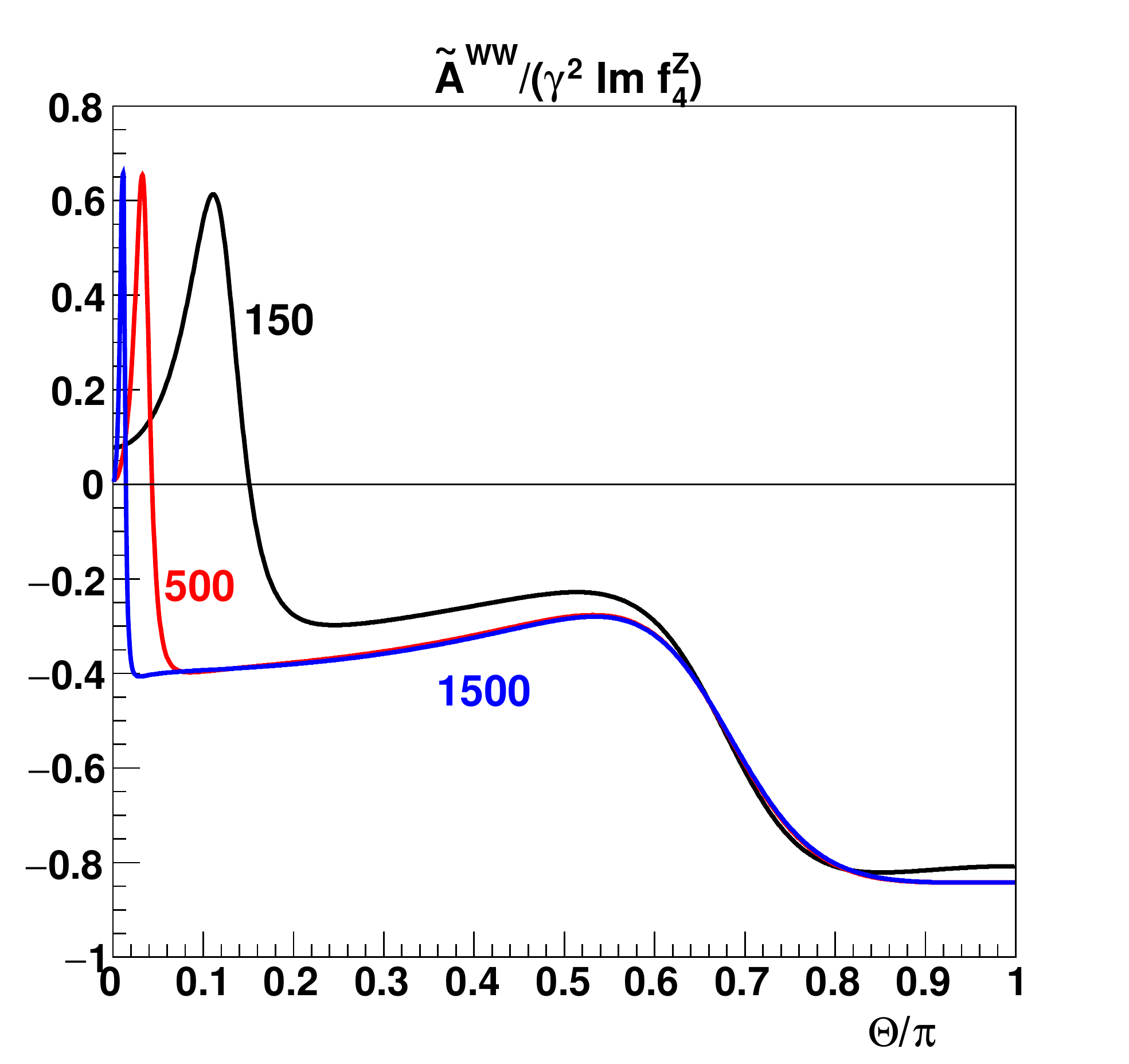}}
	\caption{The asymmetries $A^{WW}$ and $\tilde{A}^{WW}$ (divided by $\gamma^2\,\Im f_4^Z$) vs $\Theta$ for three values of the beam (or $W$) energy $E$, 150~GeV, 500~GeV and 1500~GeV, as indicated.}
	\label{fig:a_atilde_ww}
\end{figure}

\begin{figure}[htb]
	\centerline{
		\includegraphics[width=9.0cm]{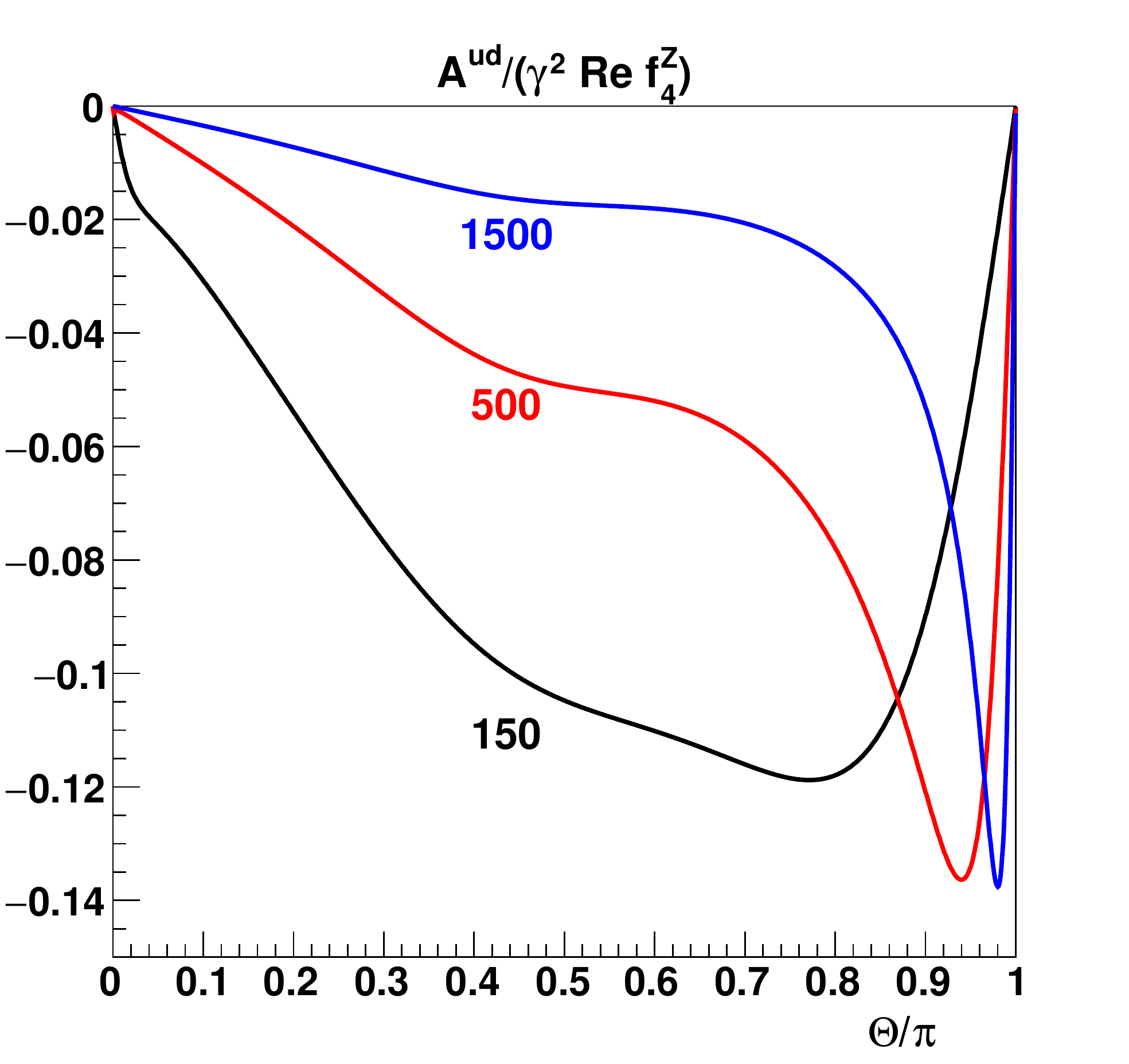}}
	\caption{The asymmetry ${\cal A}^{ud}$ of Eq.~(\ref{Eq:A_ud}) (divided by $\gamma^2\,\Re f_4^Z$) vs $\Theta$ for three values of the beam (or $W$) energy $E$, 150~GeV, 500~GeV and 1500~GeV, as indicated.}
	\label{fig:a_ud}
\end{figure}

A further possibility of testing CP violation in $e^+e^-\to WW$ has been investigated in \cite{Chang:1993vv}. Adopting the helicity amplitudes obtained there, they have defined the up-down asymmetry ${\cal A}^{ud}(\Theta)$ in their Eq.~(32), and shown that it equals
\begin{equation}
	{\cal A}^{ud}(\Theta)
	=\frac{3}{8}\sqrt{2}
	\left[\Im\rho(\Theta)_{+,0}-\Im\bar{\rho}(\Theta)_{-,0}
	-\Im\rho(\Theta)_{-,0}+\Im\bar{\rho}(\Theta)_{+,0}\right],
\end{equation}
with $\rho(\Theta)$ the spin-density matrix of the $W^-$ boson and
$\bar{\rho}(\Theta)$ the spin-density matrix of the $W^+$ boson, as
defined by their Eqs.~(26) and (28). 
To lowest order in $f_4^Z$, this quantity is proportional to $\Re f_4^Z$.
It is a rather complicated function, depending on the $W$ velocity $\beta$, the angle $\Theta$, the ratio
$M_Z^2/s_1$,
as well as $\sin^2\thetaW$. We focus on the angular dependence, and write it as
\begin{equation} \label{Eq:A_ud}
{\cal A}^{ud}
=3 \beta\sqrt{1-\beta^2} (1+\beta^2-2\beta\cos\Theta)\sin \Theta{\cal F}(s_1,\Theta)\, \Re f_4^Z,
\end{equation}
with 
\begin{equation}
{\cal F}(s_1,\Theta)
\equiv \frac{N_0^{ud}+N_1^{ud}\cos\Theta+N_2^{ud}\cos^2\Theta}
{D_0^{ud}+D_1^{ud}\cos\Theta+D_2^{ud}\cos^2\Theta+D_3^{ud}\cos^3\Theta+D_4^{ud}\cos^4\Theta}
\end{equation}
given in Appendix~D. The angular dependence of this asymmetry is shown in Fig.~\ref{fig:a_ud}.

In the low-energy limit, $\beta\to0$, this reduces to
\begin{equation}
{\cal A}^{ud} \to -\frac{3}{4} \beta \frac{(1-2 \sin^2\theta_W)M_W^2}{4M_W^2-M_Z^2} \sin \Theta\, \Re f_4^Z.
\end{equation}

On the other hand, at high energies, the prefactor ${\cal F}(s_1,\Theta)$ grows as $\gamma^2$ (see Appendix~D), but this is tempered by the high-energy fall-off of $f_4^Z$.

\begin{figure}[htb]
	\centerline{
		\includegraphics[width=8.0cm]{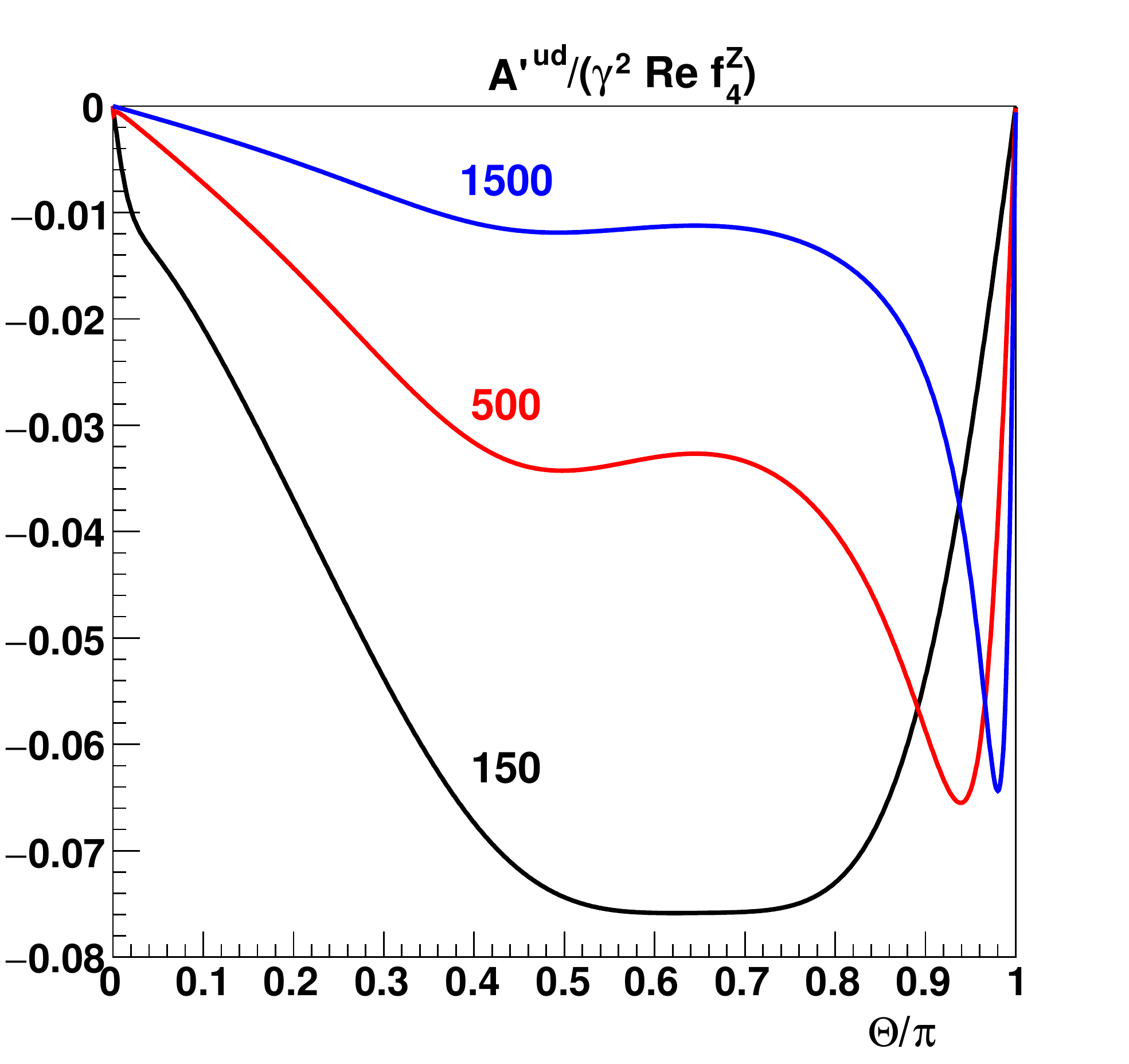}
		\includegraphics[width=8.0cm]{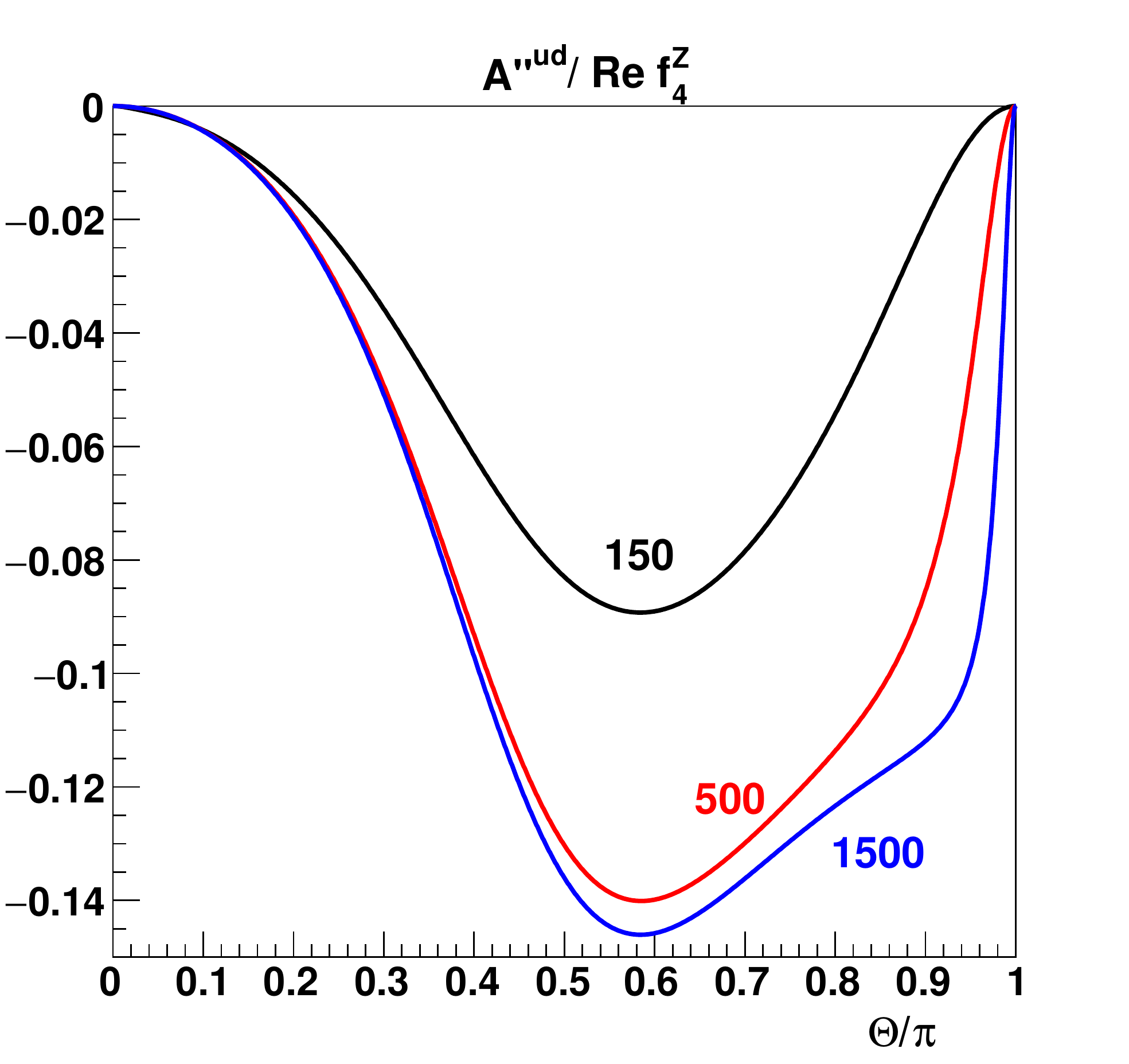}}
	\caption{The asymmetries ${\cal A}^{\prime ud}$ and ${\cal A}^{\prime \prime ud}$ divided by $\gamma^2 \Re f_4^Z$  and $\Re f_4^Z$, respectively, vs $\Theta$ for three values of the beam (or $W$) energy $E$, 150~GeV, 500~GeV and 1500~GeV, as indicated. Here, $E_0=\frac{1}{4}\sqrt{s_1}$ has been used.}
	\label{fig:ap_app_ud}
\end{figure}

Chang, Keung and Phillips \cite{Chang:1993vv} have also defined an asymmetry ${\cal A}^{\prime ud}(\Theta)$ in their Eq.~(34), and shown that it equals
\begin{eqnarray}
	{\cal A}^{\prime ud}
	&=&\frac{3\sqrt{2}}{4\pi}
	\left\{\left[a(E_0)-b(E_0)\right]
	\left[\Im\rho(\Theta)_{+,0}-\Im\bar{\rho}(\Theta)_{-,0}\right]\right.\nonumber\\
	&&\left.\hspace*{1cm}
	-\left[a(E_0)+b(E_0)\right]
	\left[\Im\rho(\Theta)_{-,0}-\Im\bar{\rho}(\Theta)_{+,0}\right]\right\},
\end{eqnarray}
with $a(E_0)$ and $b(E_0)$ defined in \cite{Chang:1993vv} following
their Eq.~(34).
To the lowest order in $f_4^Z$ we find that 
\begin{eqnarray}
	{\cal A}^{\prime ud}
	&=&\frac{3\beta\sqrt{1-\beta^2} (1+\beta^2-2\beta\cos\Theta)\sin \Theta\, s_1\, \Re f_4^Z}{\pi\left(D_0^{ud}+D_1^{ud}\cos\Theta+D_2^{ud}\cos^2\Theta+D_3^{ud}\cos^3\Theta+D_4^{ud}\cos^4\Theta\right)}\nonumber\\
	&&\times\left\{\left[a(E_0)-b(E_0)\right]{\cal N}(\beta,\cos\Theta)
	+\left[a(E_0)+b(E_0)\right]{\cal N}(-\beta,-\cos\Theta)
	\right\},
\end{eqnarray}
where
\bea
{\cal N}(\beta,\cos\Theta)&=& N_0^{\prime ud}+N_1^{\prime ud}\cos\Theta+N_2^{\prime ud}\cos^2\Theta
\eea
and
\bea
N_0^{\prime ud}&=&
\left(1-2 \beta ^2-\beta ^3\right) (1-\beta )^2 \left(1-2 \sin ^2\theta _W\right)s_1\nonumber\\
&&- \left(1-4 \beta-\beta ^2 +2 \beta ^3-2 \left(1-6 \beta-\beta ^2+\beta ^3+\beta ^5 \right) \sin ^2\theta _W\right)m_Z^2,\\
N_1^{\prime ud}&=&\left(1+3 \beta +2 \beta ^2 \right) (1-\beta )^2 \left(1-2 \sin ^2\theta _W\right)s_1\nonumber\\
&&- \left[(1+\beta )^2-2 \left(1+3 \beta+5 \beta ^2+\beta ^3-2 \beta ^4 \right) \sin ^2\theta _W\right.\nonumber\\
&&\left.\hspace*{1cm}+8\beta \left(1+\beta ^2\right) \sin ^4\theta _W\right]m_Z^2,\nonumber\\
N_2^{\prime ud}&=&2 \beta ^2  \left(1-4 \sin ^2\theta _W+8 \sin ^4\theta _W\right)m_Z^2.
\eea

Finally, they have also defined
\bea
{\cal A}^{\prime\prime ud}&=&-\frac{1}{\pi}\left(\Im \rho(\Theta)_{+,-}-\Im \bar{\rho}(\Theta)_{-,+}\right),
\eea
which to the lowest order in $f_4^Z$ equals
\bea
{\cal A}^{\prime\prime ud}&=&\frac{4 \beta  \left(1-\beta ^2\right)^2 \left(1-2 \sin ^2\theta _W\right)   \left(1+\beta ^2-2 \beta  \cos \Theta \right) \sin ^2\Theta \left(s_1-m_Z^2\right) s_1\, \Re f_4^Z}{\pi\left(
	D_0^{ud}+D_1^{ud}\cos\Theta+D_2^{ud}\cos^2\Theta+D_3^{ud}\cos^3\Theta+D_4^{ud}\cos^4\Theta
	\right)}.\nonumber\\
\eea
In the low-energy limit these become:
\bea
{\cal A}^{\prime ud}&=&-\frac{3 \beta  m_W^2 \left(2 m_W^2-m_Z^2\right) \sin \Theta \Re f_4^Z}{4 \pi  m_Z^2 \left(4 m_W^2-m_Z^2\right)}\nonumber\\
&&\times \left\{\left[a(E_0)-b(E_0)\right](1+\cos\Theta) + \left[a(E_0)+b(E_0)\right] (1-\cos\Theta)\right\}
,\\
{\cal A}^{\prime\prime ud}&=&-\frac{\beta    m_W^2 \left(2 m_W^2-m_Z^2\right) \sin ^2\Theta \Re f_4^Z}{\pi  m_Z^2 \left(4 m_W^2-m_Z^2\right)}.
\eea
The asymmetries ${\cal A}^{\prime ud}$ and ${\cal A}^{\prime \prime ud}$ are shown in Fig.~\ref{fig:ap_app_ud}. Like ${\cal A}^{ud}$, they vary rapidly near the backward direction.
\section{Discussion}
\label{Sec:Discussion}
\setcounter{equation}{0}

The mixing of CP-even and odd components of the scalar fields lead to couplings among all pairs of neutral mass eigenstates and the gauge particles, which in turn lead to loop-induced trilinear couplings among the electroweak gauge particles, $W$ and $Z$. The CP-violating part of these couplings, which we have discussed here, are all proportional to the quantity $\Im J_2$.

This quantity $\Im J_2$ is proportional to the product of couplings $e_1e_2e_3$, as well as to the product of differences of masses squared, $(M_2^2-M_1^2)(M_3^2-M_2^2)(M_1^2-M_3^2)$, see Eq.~(\ref{Eq:Im_J2}). Obviously, having all three masses different is a necessary, but not sufficient condition for CP violation. Any one of the neutral Higgs particles could be odd under CP, with the other two even. This would be reflected in a vanishing product $e_1e_2e_3$.

Properties of the Higgs boson observed at the LHC \cite{Khachatryan:2014kca,Khachatryan:2014jba,Aad:2015mxa}
match those expected 
for the SM. As has been discussed in \cite{Gunion:2002zf,Dev:2014yca,Grzadkowski:2014ada,Bernon:2015qea},
the standard nature of the Higgs boson doesn't preclude the presence 
of interesting relatively low-scale BSM physics. This
landscape is being referred to as the alignment limit. If we take the discovered 125~GeV Higgs boson to be the lightest one, $H_1$, then the alignment limit implies
\begin{equation}
e_1\to v,\quad e_2\to0, \quad e_3\to 0,
\end{equation}
where $v=246~\text{GeV}$. Hence, in this exact limit, $\Im J_2$ vanishes, and the CP-violating effects discussed in this paper, would all vanish. (Actually, also $\Im J_1$ would vanish.) In general, if we parametrize the deviation of the $H_1VV$ coupling from its SM value by $\delta$, such that $e_1=v(1-\delta)$, then one can show that  for small $\delta$
\bea
|\Im J_2| < \frac{\delta }{v^6}(M_1^2-M_2^2)(M_2^2-M_3^2)(M_3^2-M_1^2).
\eea

However, the scalar sector of the 2HDM might still offer CP violation in the alignment limit, represented by the remaining invariant $\Im J_{30}$ \cite{Grzadkowski:2014ada}. This quantity can only be accessed via measurements involving pairs of charged Higgs bosons, and is thus not easily studied.
Thus, for the near future, the best prospects for finding CP violation in the 2HDM lie probably in these trilinear vector couplings, together with some deviation from the alignment limit \cite{Basso:2012st,Basso:2013wna,Fontes:2015mea,Fontes:2015xva}.

These effects of CP violation might be worth looking for also at the LHC, where in Drell--Yan processes a cut on rapidity makes it possible to statistically distinguish the quark from the antiquark direction \cite{Dittmar:1996my}. An efficiency study might be worthwhile.

Of course, if an effect were to be found in $ZZZ$ or $ZW^+W^-$ trilinear couplings, at a level beyond that expected in the 2HDM, that would point to some other new physics.

\section*{Acknowledgments}
It is a pleasure to thank professor W. Y. Keung and professor P. B. Pal for correspondence.
BG thanks Jan Kalinowski for clarifying discussions and also acknowledges partial support by the National Science Centre (Poland) research project, 
decision no DEC-2014/13/B/ST2/03969. PO is supported in part by the Research Council of Norway.

\appendix
\section{The \boldmath{$ZZZ$ vertex}}
\label{Sec:zzz}
\setcounter{equation}{0}

\subsection{The \boldmath{$HHH$} triangle diagram}
\label{Sec:zzz-hhh}
\begin{figure}[htb]
\centerline{
\includegraphics[width=7.0cm]{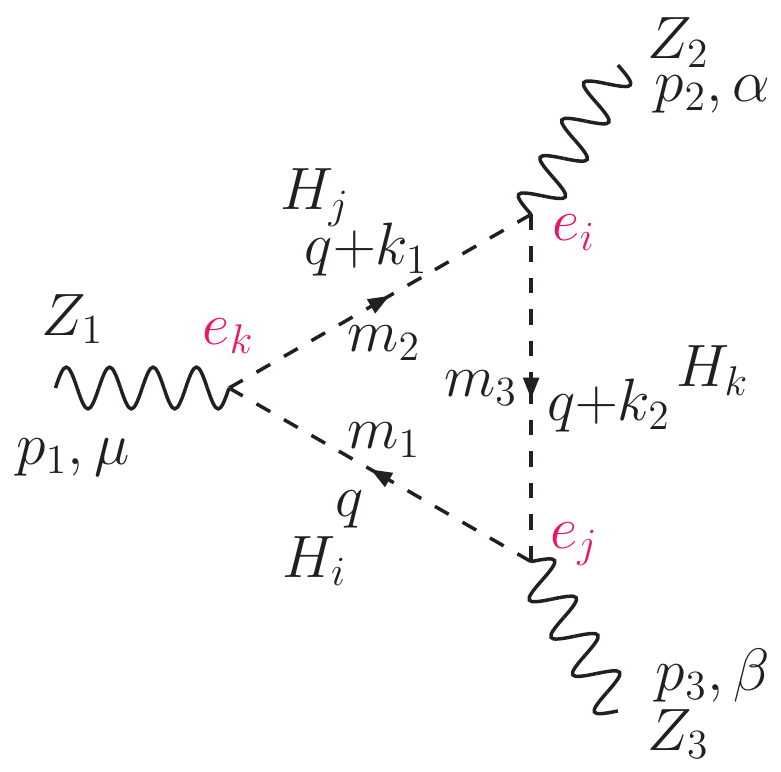}}
\caption{Triangle diagram contributing to the CP-violating $ZZZ$ vertex.}
\label{fig:Feynman-j2-a-app}
\end{figure}

We show in Fig.~\ref{fig:Feynman-j2-a-app} the triangle diagram in LoopTools notation \cite{Hahn:1998yk}. Treating all momenta $p_1$, $(-p_2)$ and $(-p_3)$ as incoming: $p_1-p_2-p_3=0$. Loop momenta along the three internal lines are denoted $q$, $q+k_1$, $q+k_2$, whereas their masses are denoted $m_1$, $m_2$ and $m_3$ (some permutation of $M_1,M_2,M_3$). 

Assuming that $Z$ couples to light fermions we may drop terms proportional to $p_1^\mu$, $p_2^\alpha$ and $p_3^\beta$. Furthermore, we assume that $Z_2$ and $Z_3$ are on-shell, meaning $p_2^2=p_3^2=M_Z^2$.
Under these assumptions, the contribution to $f_4^Z$ is given by the following sum over 6 permutations of $i,j,k$:
\begin{equation}
\label{eq:case}
e\frac{p_1^2-M_Z^2}{M_Z^2} f_4^{Z,HHH} 
= -8N_H e_1e_2e_3\sum_{i,j,k}\epsilon_{ijk}C_{001}(p_1^2,M_Z^2,M_Z^2,M_i^2,M_j^2,M_k^2),
 \end{equation}
where
\begin{equation}
N_H=\frac{1}{16\pi^2}\left(\frac{g}{2v\cos\thetaW}
\right)^3
=\frac{e\alpha}{4\pi v^3\sin^3(2\thetaW)}.
\end{equation}
 
\subsection{The \boldmath{$HHG$} triangle diagrams}
\label{Sec:zzz-hhg}
\begin{figure}[htb]
\centerline{
\includegraphics[width=7.0cm]{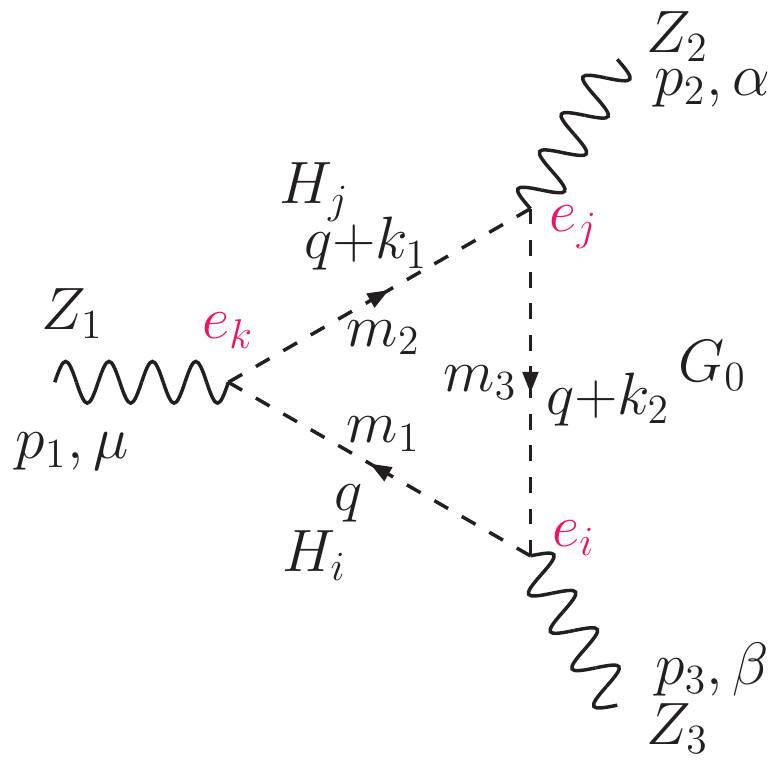}}
\caption{Triangle diagram contributing to the CP-violating $ZZZ$ vertex.}
\label{fig:Feynman-j2-ag-app}
\end{figure}

In the covariant gauge, there are also contributions from triangle diagrams with one of the Higgs fields replaced by the Goldstone field $G_0$. A representative case is shown in Fig.~\ref{fig:Feynman-j2-ag-app}. There are similar diagrams with a $G_0$ along either of the other two internal lines, but no diagram with two or three internal $G_0$ lines due to the non-existence of a $ZG_0G_0$ coupling.

We add these three sets of diagrams, summing over permutations of $i,j,k$ and make the same assumptions as for the $HHH$ triangle diagrams. We find that under these assumptions the remaining contribution to $f_4^Z$ is given by
\begin{align}\label{eq:sum-HHG}
e\frac{p_1^2-M_Z^2}{M_Z^2} f_4^{Z,HHG} 
&=8N_H e_1e_2e_3\sum_{i,j,k}\epsilon_{ijk}\bigl[
  C_{001}(p_1^2,M_Z^2,M_Z^2,M_i^2,M_j^2,M_Z^2)  \\
& +C_{001}(p_1^2,M_Z^2,M_Z^2,M_Z^2,M_j^2,M_k^2)
 +C_{001}(p_1^2,M_Z^2,M_Z^2,M_i^2,M_Z^2,M_k^2)\bigr].
 \nonumber
 \end{align}

    \subsection{The \boldmath{$HHZ$}  triangle diagrams}
\label{Sec:zzz-hhz}
\begin{figure}[htb]
\centerline{
\includegraphics[width=7.0cm]{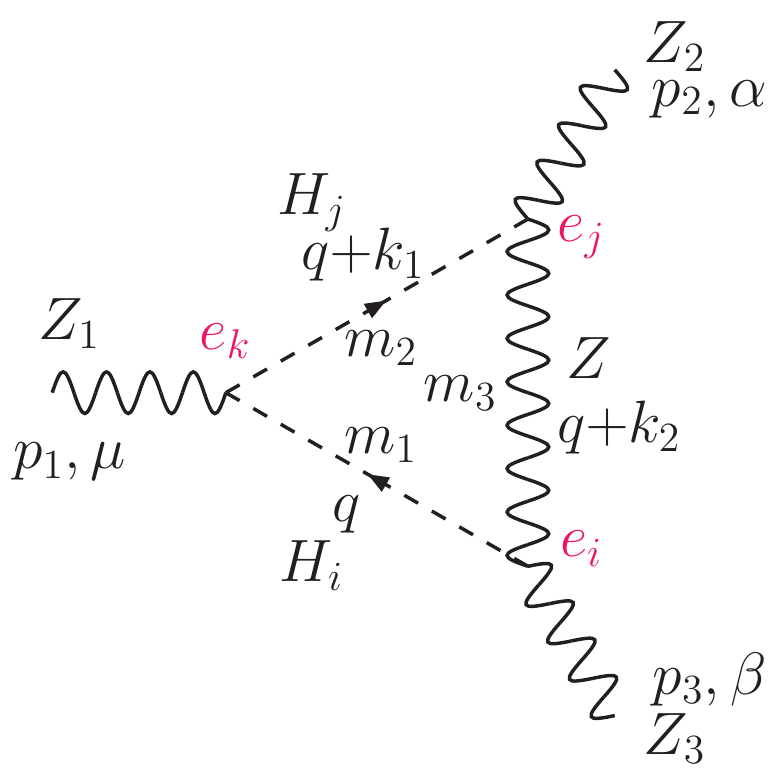}}
\caption{Triangle diagram contributing to the CP-violating $ZZZ$ vertex.}
\label{fig:Feynman-j2-az-app}
\end{figure}

One of the lines in the triangle diagram could also be a $Z$, as indicated in Fig.~\ref{fig:Feynman-j2-az-app}. The $Z$-line can of course also be inserted between $Z_3$ and $Z_1$, or between $Z_1$ and $Z_2$, but there can not be more than one internal $Z$-line due to the absence of a tree-level $ZZZ$ vertex.
Again, we add these three sets of diagrams, summing over permutations of $i,j,k$ and make the same assumptions as for the $HHH$ and $HHG$ triangle diagrams. We find that under these assumptions the remaining contribution to $f_4^Z$ is given by

\begin{align}
e\frac{p_1^2-M_Z^2}{M_Z^2} f_4^{Z,HHZ} 
&=-8M_Z^2N_H e_1e_2e_3\sum_{i,j,k}\epsilon_{ijk}
C_1(p_1^2,M_Z^2,M_Z^2,M_i^2,M_Z^2,M_k^2).
\end{align}

\subsection{Bubble diagrams}

\begin{figure}[htb]
\centerline{
\includegraphics[width=7.5cm]{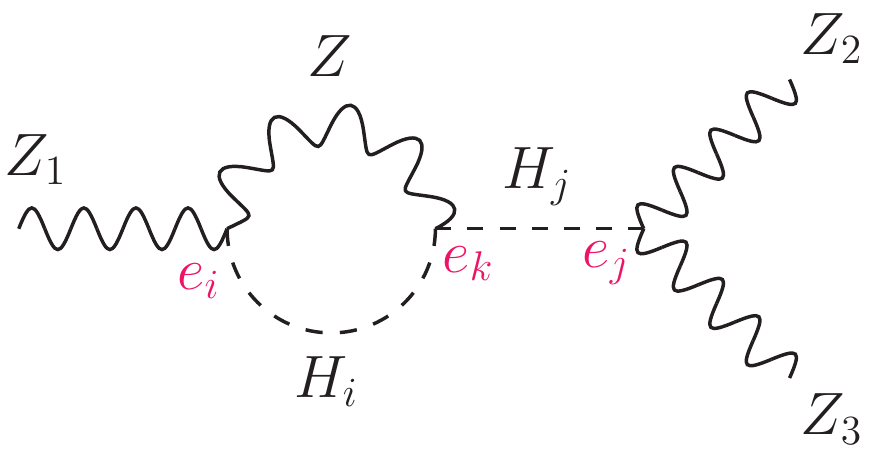}}
\caption{Bubble diagram contributing proportional to $\Im J_2$.}
\label{fig:Feynman-j2-c}
\end{figure}

There are diagrams with a bubble connecting a $Z_a$ ($a=1,2,3$) with an intermediate $H_j$, as shown in Fig.~\ref{fig:Feynman-j2-c}.

These diagrams contribute terms proportional to $\Im J_2$. There are also diagrams with the internal $Z$ replaced by a $G_0$. However, these diagrams are all scalar, proportional to $p_1^{\mu}g^{\alpha\beta}$ (and similarly for bubbles on the other legs) and will not be further discussed.

\begin{figure}[htb]
\centerline{
\includegraphics[width=5.5cm]{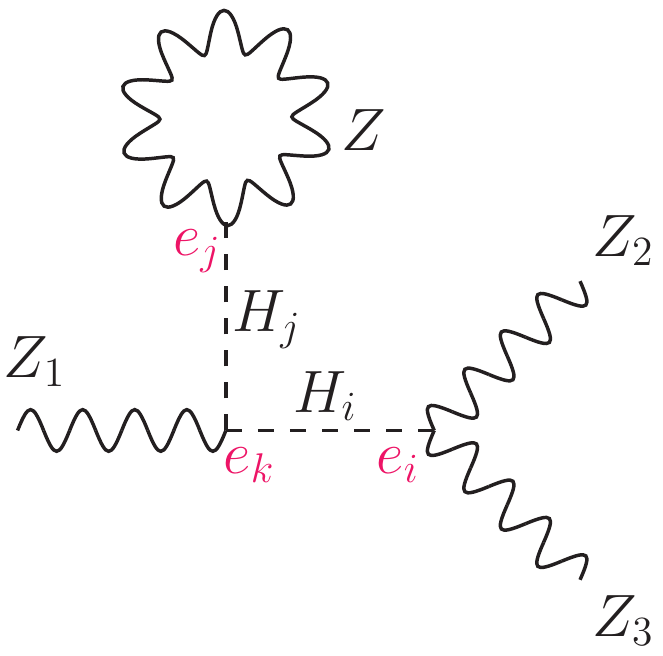}}
\caption{Tadpole diagram yielding a structure proportional to $\Im J_2$.}
\label{fig:Feynman-j2-b}
\end{figure}

\subsection{Tadpole diagrams}

There are also tadpole diagrams yielding a structure proportional to $\Im J_2$. A representative case is shown in Fig.~\ref{fig:Feynman-j2-b}. However, these are canceled by counter terms, in order to give a vanishing expectation value for the $H_j$ field at the one-loop order \cite{Santos:1996vt}.

\section{The \boldmath{$ZW^+W^-$ vertex}}
\label{Sec:zww}
\setcounter{equation}{0}

\begin{figure}[htb]
\centerline{
\includegraphics[width=7.0cm]{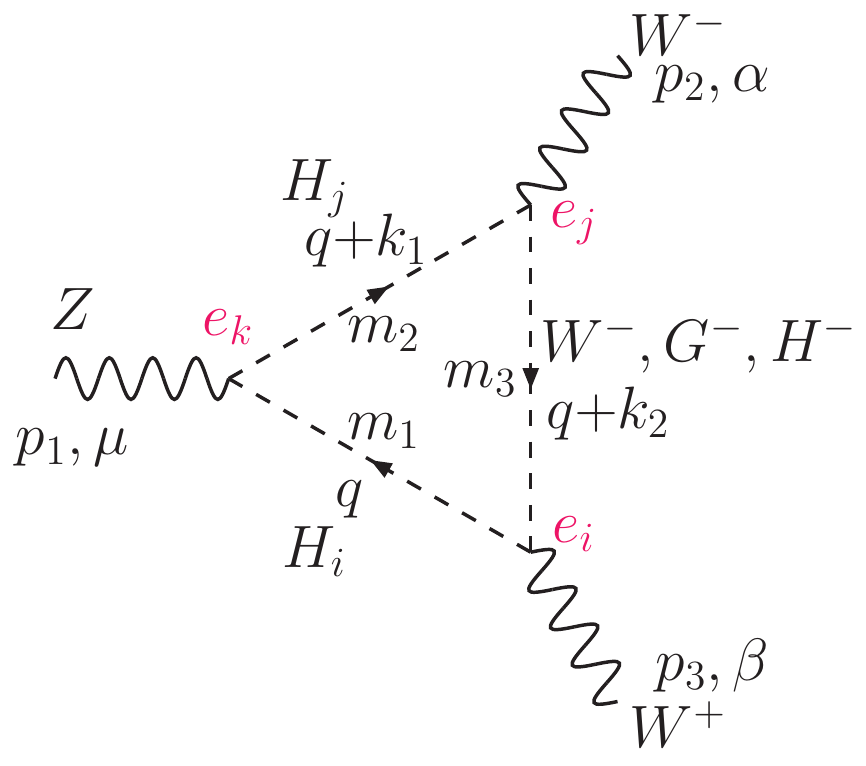}}
\caption{Triangle diagrams contributing to the CP-violating $ZWW$ vertex.}
\label{fig:Feynman-j2-ZWW}
\end{figure}

Triangle diagrams of the kind shown in Fig.~\ref{fig:Feynman-j2-ZWW}
contribute to the CP-violating $ZW^+W^-$ vertex. In fact, they give a
contribution proportional to the invariant $\Im J_2$.

We show in Fig.~\ref{fig:Feynman-j2-ZWW} the triangle diagrams contributing to the CP-violating form factor in LoopTools notation \cite{Hahn:1998yk}. The details of their calculations and the assumptions made are similar to the calculations of the $ZZZ$ vertex in the previous section, so we omit the details.

For those diagrams with a $W^-$ line between $W^+$ and $W^-$, we find that their contribution is proportional to $p_1^{\mu}g^{\alpha\beta}$. We therefore neglect this contribution.
Putting
\begin{equation}
	N=\frac{-1}{16\pi^2\cos\thetaW}\left(\frac{g}{2v}\right)^3
	=\frac{-e\alpha}{32\pi v^3\cos\thetaW\sin^3(\thetaW)},
\end{equation}
we find that
for the diagrams with  a $G^-$ line between $W^+$ and $W^-$, their contribution is
\begin{align}
	ig_{ZWW}\Gamma_{ZWW,HHG_{\rm ch}}^{\alpha\beta\mu}
	=8N e_1e_2e_3(p_1^{\alpha}g^{\mu\beta}+p_1^{\beta}g^{\mu\alpha})\sum_{i,j,k}\epsilon_{ijk}
	C_{001}(p_1^2,M_W^2,M_W^2,M_i^2,M_j^2,M_W^2).
\end{align}

As for the diagrams with an $H^-$ line between $W^+$ and $W^-$, there are contributions to the CP-violating form factor as well as to CP-conserving ones. We present only the contribution to the CP-violating form factor, which becomes
\begin{align}
	ig_{ZWW}\Gamma_{ZWW,HHH_{\rm ch}}^{\alpha\beta\mu}
	&=-8N e_1e_2e_3(p_1^{\alpha}g^{\mu\beta}+p_1^{\beta}g^{\mu\alpha})\sum_{i,j,k}\epsilon_{ijk}
	C_{001}(p_1^2,M_W^2,M_W^2,M_i^2,M_j^2,M_{H^\pm}^2)\nonumber\\
	&+{\text{CP-conserving terms}}.
\end{align}

Furthermore, at the SM level, there are contributions from fermion loops.
But these do not contribute to $f_4^Z$ and will be ignored.
\section{Extracting \boldmath{$\Im J_2$} --- a case study}
\label{Sec:extract}
\setcounter{equation}{0}
We have claimed in Sections \ref{Sec:ZZZ} and \ref{Sec:ZWW} that the contributions to $f_4^Z$ from the diagrams studied in Appendices \ref{Sec:zzz} and \ref{Sec:zww} are proportional to $\Im J_2$. However, $\Im J_2$ is not explicit in the expressions (\ref{eq:f4-zzz}) and (\ref{eq:f_4W}). The expressions contain the factor $e_1 e_2 e_3$, but the factor $(M_1^2-M_2^2)(M_2^2-M_3^2)(M_3^2-M_1^2)$ is not explicitly visible. This remaining factor is ``hidden'' in the linear combination of C-functions. One readily finds numerically that the expressions for $f_4^Z$ vanish whenever two of the scalars have the same mass, but extracting the ``hidden'' factor $(M_1^2-M_2^2)(M_2^2-M_3^2)(M_3^2-M_1^2)$ analytically is not easily done. Making some assumptions makes this task easier.

Let us study the contribution from the triangle diagrams with $H_iH_jH_k$ in the loop. The contribution is presented in (\ref{eq:case}). Let us focus on the expression
\bea
\Sigma=\sum_{i,j,k}\epsilon_{ijk}C_{001}(p_1^2,M_Z^2,M_Z^2,M_i^2,M_j^2,M_k^2)
\eea
in the asymptotic limit (large $s_1\equiv p_1^2$). The tensor coefficients $C_{001}$ can all be re-expressed in terms of scalar loop integrals $A_0$, $B_0$ and $C_0$ \cite{De:2003}. The explicit forms of these scalar loop integrals are all known \cite{De:1993}. They can be expressed in terms of logarithms and dilogarithms. The resulting expression is very lengthy and complex, so we prepare to study the expression in the asymptotic limit by introducing new variables:
\begin{eqnarray}
x&=&\frac{(M_2^2-M_1^2)}{s_1}\\
y&=&\frac{(M_3^2-M_2^2)}{s_1}
\eea
As a consequence,
\bea
x+y&=&\frac{(M_3^2-M_1^2)}{s_1},
\end{eqnarray}
and
\begin{eqnarray}
\Im J_2&=&2\frac{e_1 e_2 e_3}{v^9}s_1^3 xy(x+y).\nonumber
\end{eqnarray}

Both $x$ and $y$ are small in the asymptotic limit. Expanding $\Sigma$ in power of $x$ and $y$, we find that the leading term indeed contains the product $xy(x+y)$,
\begin{eqnarray}
f_4^{Z,HHH}
&=&\frac{-2\alpha}{\pi\sin^3(2\thetaW)}\frac{M_Z^2}{p_1^2-M_Z^2}
\frac{e_1e_2e_3}{v^3}
\sum_{i,j,k}\epsilon_{ijk}
C_{001}(p_1^2,M_Z^2,M_Z^2,M_i^2,M_j^2,M_k^2)  \nonumber\\
&\simeq&\frac{-\alpha}{4\pi\sin^3(2\thetaW)}\frac{v^6M_Z^2}{M_1^2 s_1^2(s_1-M_Z^2)}
{\rm Im} J_2\nonumber\\
&&\times
\left(\log \left(\frac{M_1^2}{s_1}\right)+\frac{i \left(9 M_1^2-2 M_Z^2\right) \log \left(\frac{\sqrt{4 M_1^2-M_Z^2}-i M_Z}{\sqrt{4 M_1^2-M_Z^2}+i M_Z}\right)}{M_Z \sqrt{4 M_1^2-M_Z^2}}+i \pi \right)
\nonumber
\end{eqnarray}
in the asymptotic limit. Here, both $M_2^2$ and $M_3^2$ have been represented by $M_1^2$, since $x\ll1$ and $y\ll1$.

\section{Some asymmetry prefactors \boldmath{${\cal F}$}}
\label{Sec:prefactor}
\setcounter{equation}{0}

\subsection{The prefactor \boldmath{${\cal F}_1(\beta,\Theta)$} of \boldmath{${\cal A}_1^{ZZ}$}}
	
We define the prefactor of Eq.~(\ref{a1zz}) as
\begin{equation}
{\cal F}_1(\beta,\Theta)
=\frac{N_0+N_1\cos\Theta+N_2\cos^2\Theta+N_3\cos^3\Theta}
{D_0+D_1\cos \Theta+D_2\cos^2 \Theta+D_3\cos^3 \Theta+D_4\cos^4 \Theta}.
\end{equation}
These coefficients are given by
\begin{subequations}
\begin{alignat}{2}
N_0&=\left(1+\beta ^2\right) \xi _1, &\quad
N_1&=-2 \beta ^2 \left(\xi _1-\xi _2\right),\\
N_2&=\left(\beta ^2-3\right) \xi _1, &\quad
N_3&=2 \left(\xi _1-\xi _2\right),\\
D_0&=\left(1+\beta ^2\right)^2 \left(\xi _3+\xi _4\right), &\quad
D_1&=2 \left(1-\beta ^4\right) \xi _3,\\
D_2&=-\left(3+6 \beta ^2-\beta ^4\right) \left(\xi _3+\xi _4\right), &\quad
D_3&=-4 \left(1-\beta ^2\right) \xi _3,\\
D_4&=4 \left(\xi _3+\xi _4\right),
\end{alignat}
\end{subequations}
with
\begin{subequations}
\begin{align}
\xi_1&=\sin\theta_W\cos\theta_W(1-6\sin^2\theta_W+12\sin^4\theta_W),\\
\xi_2&=16\sin^7\theta_W\cos\theta_W,\\
\xi_3&=1-8\sin^2\theta_W+24\sin^4\theta_W-32\sin^6\theta_W,\\
\xi_4&=32\sin^8\theta_W.
\end{align}
\end{subequations}

\subsection{The prefactor \boldmath{${\cal F}(s_1,\Theta)$} of \boldmath{${\cal A}^{ud}$}}
The prefactor of the asymmetry ${\cal A}^{ud}$ of Eq.~(\ref{Eq:A_ud}) can be written as
\begin{equation}
{\cal F}(s_1,\Theta)=\frac{N_0^{ud}+N_1^{ud}\cos\Theta+N_2^{ud}\cos^2\Theta}{D_0^{ud}+D_1^{ud}\cos\Theta+D_2^{ud}\cos^2\Theta+D_3^{ud}\cos^3\Theta+D_4^{ud}\cos^4\Theta}
\end{equation}
with
\begin{subequations}
\begin{align}
N_0^{ud}&=\left(1-\beta ^2\right)  \left(1-2 \sin ^2\theta _W\right)\left(s_1-m_Z^2\right),\\
N_1^{ud}&=\beta  \left[\left(1-\beta ^2\right) \left(1-2 \sin ^2\theta _W\right)s_1\right.\nonumber\\
&\left.\hspace*{0.5cm}- \left(2-2 \left(3+\beta ^2\right) \sin ^2\theta _W+8 \left(1+\beta ^2\right) \sin ^4\theta _W\right)m_Z^2\right],\\
N_2^{ud}&=2 \beta ^2 \left(1-4 \sin ^2\theta _W+8 \sin ^4\theta _W\right) m_Z^2,
\end{align}
\end{subequations}
and
\begin{subequations}
\begin{align}
D_0^{ud}&=-\left(1-\beta ^2\right)^2 \left(16+11 \beta ^2-18 \beta ^4+3 \beta ^6\right) s_1^2\nonumber\\
&+4 \left(1-\beta ^2\right)  \left(8+7 \beta ^2-14 \beta ^4+3 \beta ^6- \beta ^2\left(19+\beta ^2-15 \beta ^4+3 \beta ^6\right) \sin ^2\theta _W\right) m_Z^2 s_1\nonumber\\
&-4  \left(4+9 \beta ^2-10 \beta ^4+\beta ^6-2\beta ^2 \left(19+\beta ^2-15 \beta ^4+3 \beta ^6\right)  \sin ^2\theta _W\right.\nonumber\\
&\left.\hspace*{1cm}+2 \beta^2\left(1+\beta ^2 \right)^2 \left(19-18 \beta ^2+3 \beta ^4\right) \sin ^4\theta _W\right)m_Z^4,\\
D_1^{ud}&=4 \beta  \left[\left(1-\beta ^2\right)^2 \left(8-9 \beta ^2+3 \beta ^4\right) s_1^2\right.\nonumber\\
&\hspace*{0.7cm}-4 \left(1-\beta ^2\right) \left(6-4 \beta ^2-\left(4+9 \beta ^2-12 \beta ^4+3 \beta ^6\right) \sin ^2\theta _W\right) m_Z^2  s_1\nonumber\\
&\hspace*{0.7cm}+4 \left(4+\beta ^2-3 \beta ^4-4 \left(1+6 \beta ^2-5 \beta ^4\right) \sin ^2\theta _W\right.\nonumber\\
&\left.\left.\hspace*{1.4cm}+2 \beta ^2 \left(19+\beta ^2-15 \beta ^4+3 \beta ^6\right)  \sin ^4\theta _W\right) m_Z^4\right],\\
D_2^{ud}&=\beta ^2 \left(1-\beta ^2\right) \left[\left(1-\beta ^2\right)^2 \left(11-3 \beta ^2\right) s_1^2\right.\nonumber\\
&\hspace*{3cm}+4 \left(1-\beta ^2\right)  \left(3 \left(3+\beta ^2\right)-\left(29-8 \beta ^2+3 \beta ^4\right) \sin ^2\theta _W\right) m_Z^2 s_1\nonumber\\
&\hspace*{3cm}-4  \left(11+5 \beta ^2-\left(26+32 \beta ^2+6 \beta ^4\right) \sin ^2\theta _W\right.\nonumber\\
&\hspace*{4cm}\left.\left.- \left(6-138 \beta ^2+10 \beta ^4-6 \beta ^6\right) \sin ^4\theta _W\right)m_Z^4
\right],\\
D_3^{ud}&=4 \beta ^3 \left[\left(1-\beta ^2\right)^2 \left(1-3 \beta ^2\right) s_1^2+4 \left(1-\beta ^2\right) \left(2 \beta ^2-\left(1+3 \beta ^4\right) \sin ^2\theta _W\right) m_Z^2 s_1\right.\nonumber\\
&\hspace*{2cm}-4 \left(1+\beta ^2-4 \left(1+\beta ^4\right) \sin ^2\theta _W\right.\nonumber\\
&\left.\left.\hspace*{3cm}+\left(6+2\beta ^2+2\beta ^4+6 \beta ^6\right) \sin ^4\theta _W\right)  m_Z^4 \right],\\
D_4^{ud}&=4 \beta ^4 \left(3 \left(1-\beta ^2\right)^2 s_1^2-4 \left(1-\beta ^2\right) \left(1+\left(1-3 \beta ^2\right) \sin ^2\theta _W\right) m_Z^2 s_1\right.\nonumber\\
&\left.\hspace*{1cm}+4  \left(1-2 \left(1+\beta ^2\right) \sin ^2\theta _W+\left(6-4 \beta ^2+6 \beta ^4\right) \sin ^4\theta _W\right)m_Z^4\right).
\end{align}
\end{subequations}
At high energies ($s_1\gg M_Z^2$), the prefactor grows as $\gamma^2$. In the perpendicular direction, $\cos\Theta\to0$, it takes the form
\begin{equation}
{\cal F}(s_1,\Theta) \to
-\frac{\gamma^2 M_W^4(2M_W^2-M_Z^2)}{M_Z^2(12M_W^4-8 M_W^2 M_Z^2 +5 M_Z^4)},
\end{equation}
whereas in the forward direction it is more singular. That singularity is however tamed by the other factors of Eq.~(\ref{Eq:A_ud}).

\subsection{The prefactors \boldmath{${\cal F}^{WW}$} and \boldmath{${\cal \tilde F}^{WW}$} of \boldmath{$A^{WW}$} and \boldmath{$\tilde A^{WW}$}}
We define
\begin{equation}
{\cal F}^{WW}
\equiv \frac{(N_0+N_1\cos\Theta+N_2\cos^2\Theta)s_1}
	{D_0+D_1\cos\Theta+D_2\cos^2\Theta+D_3\cos^3\Theta+D_4\cos^4\Theta},
\end{equation}
and
\begin{equation}
{\cal \tilde F}^{WW}
\equiv 
\frac{(\tilde{N}_0+\tilde{N}_1\cos\Theta+\tilde{N}_2\cos^2\Theta+\tilde{N}_3\cos^3\Theta)s_1}
	{D_0+D_1\cos\Theta+D_2\cos^2\Theta+D_3\cos^3\Theta+D_4\cos^4\Theta}.
\end{equation}
The coefficients are given by
\begin{subequations}
\begin{align}
N_0&=\left(1-\beta ^2\right) \left(1-2 \sin ^2\theta _W\right) \left(s_1-m_Z^2\right) ,\\
N_1&=2 \beta  \left[\left(1-\beta ^2\right) \left(1-2 \sin ^2\theta _W\right)s_1-2  \left(1-\left(3+\beta ^2\right) \sin ^2\theta _W\right)m_Z^2\right],\\
N_2&=-3 \left(1-\beta ^2\right) \left(1-2 \sin ^2\theta _W\right)s_1+ \left(3+\beta ^2- \left(6+10 \beta ^2\right) \sin ^2\theta _W\right)m_Z^2,
\end{align}
\end{subequations}
\begin{subequations}
\begin{align}
\tilde{N}_0&=-\beta  \left(1-\beta ^2\right) \left(1-2 \sin ^2\theta _W\right)s_1\nonumber\\
&+2 \beta  \left(1-\left(3+\beta ^2\right) \sin ^2\theta _W+4 \left(1+\beta ^2\right) \sin ^4\theta _W\right) m_Z^2,\\
\tilde{N}_1&=-2 \beta ^2  \left(1-4 \sin ^2\theta _W+8 \sin ^4\theta _W\right) m_Z^2,\\
\tilde{N}_2&=-\beta  \left[\left(1-\beta ^2\right)  \left(1-2 \sin ^2\theta _W\right)s_1\right.\nonumber\\
&\left.\hspace*{0.8cm}- \left(2-2 \left(3+\beta ^2\right) \sin ^2\theta _W+8 \left(1+\beta ^2\right) \sin ^4\theta _W\right)m_Z^2\right],\\
\tilde{N}_3&=2\left(1-\beta ^2\right) \left(1-2 \sin ^2\theta _W\right)s_1-2 \left(1-2 \left(1+\beta ^2\right) \sin ^2\theta _W+8 \beta ^2 \sin ^4\theta _W\right)m_Z^2,
\end{align}
\end{subequations}
and
\begin{subequations}
\begin{align}
D_0&=\left(1+\beta ^2\right) \left[\left(1-\beta ^2\right)^2 s_1^2-2 \left(1-\beta ^2\right)  \left(1-2 \beta ^2 \sin ^2\theta _W\right) m_Z^2 s_1\right.\nonumber\\
&\left.\hspace*{1.7cm}+ \left(1+\beta ^2-8 \beta ^2 \sin ^2\theta _W+8\beta^2 \left(1+\beta ^2\right) \sin ^4\theta _W\right)m_Z^4\right],\\
D_1&=4 \beta ^3 m_Z^2 \left[\left(1-\beta ^2\right) \left(1-2 \sin ^2\theta _W\right)s_1\right.\nonumber\\
&\left.\hspace*{1.5cm}- \left(2-2 \left(3+\beta ^2\right) \sin ^2\theta _W+8 \left(1+\beta ^2\right) \sin ^4\theta _W\right)m_Z^2\right],\\
D_2&=-\left(3-\beta ^2\right) \left(1-\beta ^2\right)^2 s_1^2+2  \left(3-8 \beta ^2+5 \beta ^4+2\beta ^2 \left(1-\beta ^4\right)  \sin ^2\theta _W\right) m_Z^2 s_1\nonumber\\
&- \left(3-10 \beta ^2-\beta ^4+8\beta^2 \left(1+3 \beta ^2\right) \sin ^2\theta _W-8\beta^2 \left(1+6 \beta ^2+\beta ^4\right) \sin ^4\theta _W\right)m_Z^4, \\
D_3&=-4 \beta  \left[\left(1-\beta ^2\right)^2 s_1^2
-\left(1-\beta ^2\right)  \left(3-\left(2+4 \beta ^2\right) \sin ^2\theta _W\right)m_Z^2 s_1\right.\nonumber\\
&\left.\hspace*{1cm} +\left(2-2 \left(1+3 \beta ^2\right) \sin ^2\theta _W+8\beta^2 \left(1+\beta ^2\right) \sin ^4\theta _W\right)m_Z^4\right],\\
D_4&=4\left(1-\beta ^2\right)^2 s_1^2 -8 \left(1-\beta ^2\right) \left(1-2 \beta ^2 \sin ^2\theta _W\right) m_Z^2 s_1\nonumber\\
&+4 \left(1-4 \beta ^2 \sin ^2\theta _W+8 \beta ^4 \sin ^4\theta _W\right)m_Z^4.
\end{align}
\end{subequations}

\bibliography{paper_3}
\bibliographystyle{jhep}

\end{document}